\documentclass[english,journal=jctcce,manuscript=article,articletitle=true,layout=onecolumn]{achemso}
\pdfoutput=1
\usepackage[T1]{fontenc}
\usepackage[utf8]{inputenc}
\usepackage{babel}
\usepackage{verbatim}
\usepackage{refstyle}
\usepackage{amsmath}
\usepackage{graphicx}
\usepackage{rotfloat}
\usepackage[numbers]{natbib}
\usepackage[unicode=true,pdfusetitle,
 bookmarks=true,bookmarksnumbered=false,bookmarksopen=false,
 breaklinks=false,pdfborder={0 0 1},backref=false,colorlinks=false]
 {hyperref}

\makeatletter

\title{Straightforward and accurate automatic auxiliary basis set generation
for molecular calculations with atomic orbital basis sets}

\author{Susi Lehtola}

\email{susi.lehtola@alumni.helsinki.fi}

\affiliation{Molecular Sciences Software Institute, Blacksburg, Virginia 24061,
United States}

\AtBeginDocument{\providecommand\eqref[1]{\ref{eq:#1}}}
\AtBeginDocument{\providecommand\secref[1]{\ref{sec:#1}}}
\AtBeginDocument{\providecommand\algref[1]{\ref{alg:#1}}}
\AtBeginDocument{\providecommand\tabref[1]{\ref{tab:#1}}}
\AtBeginDocument{\providecommand\subsecref[1]{\ref{subsec:#1}}}
\AtBeginDocument{\providecommand\figref[1]{\ref{fig:#1}}}
\providecommand{\tabularnewline}{\\}
\floatstyle{ruled}
\newfloat{algorithm}{tbp}{loa}
\providecommand{\algorithmname}{Algorithm}
\floatname{algorithm}{\protect\algorithmname}
\RS@ifundefined{subsecref}
  {\newref{subsec}{name = \RSsectxt}}
  {}
\RS@ifundefined{thmref}
  {\def\RSthmtxt{theorem~}\newref{thm}{name = \RSthmtxt}}
  {}
\RS@ifundefined{lemref}
  {\def\RSlemtxt{lemma~}\newref{lem}{name = \RSlemtxt}}
  {}

\SectionNumbersOn
\usepackage[version=4]{mhchem}

\@ifundefined{showcaptionsetup}{}{%
 \PassOptionsToPackage{caption=false}{subfig}}
\usepackage{subfig}
\makeatother

\begin{document}
\begin{abstract}
Density fitting (DF), also known as the resolution of the identity
(RI), is a widely used technique in quantum chemical calculations
with various types of atomic basis sets---Gaussian-type orbitals,
Slater-type orbitals, as well as numerical atomic orbitals---to speed
up density functional, Hartree--Fock, and post-Hartree--Fock calculations.
Traditionally, custom auxiliary basis sets are hand-optimized for
each orbital basis set; however, some automatic schemes have also
been suggested. In this work, we propose a simple yet numerically
stable automated scheme for forming auxiliary basis sets with the
help of a pivoted Cholesky decomposition, which is applicable to any
type of atomic basis function. We exemplify the scheme with proof-of-concept
calculations with Gaussian basis sets and show that the proposed approach
leads to negligible DF/RI errors in Hartree--Fock (HF) and second-order
Møller--Plesset (MP2) total energies of the non-multireference part
of the W4-17 test set when used with orbital basis sets of at least
polarized triple-$\zeta$ quality. The results are promising for future
applications employing Slater-type orbitals or numerical atomic orbitals,
as well as schemes based on the use of local fitting approaches. Global
fitting approaches can also be used, in which case the high-angular-momentum
functions produced by the present scheme can be truncated to minimize
the computational cost.
\end{abstract}
\newcommand*\ie{{\em i.e.}}
\newcommand*\eg{{\em e.g.}}
\newcommand*\etal{{\em et al.}}
\newcommand*\citeref[1]{ref. \citenum{#1}}
\newcommand*\citerefs[1]{refs. \citenum{#1}} 

\let\algref=\relax
\newref{alg}{name = algorithm~, names = algorithms~, Name = Algorithm~, Names = Algorithms~}

\newcommand*\Erkale{{\sc Erkale}}
\newcommand*\Bagel{{\sc Bagel}}
\newcommand*\FHIaims{{\sc FHI-aims}}
\newcommand*\LibXC{{\sc LibXC}}
\newcommand*\Orca{{\sc Orca}}
\newcommand*\PySCF{{\sc PySCF}}
\newcommand*\PsiFour{{\sc Psi4}}
\newcommand*\Turbomole{{\sc Turbomole}}

\section{Introduction \label{sec:Introduction}}

Density fitting (DF),\citep{Whitten1973_JCP_4496,Baerends1973_CP_41,Dunlap1977_IJQC_81,Dunlap1979_JCP_3396,Dunlap2010_MP_3167}
also known as the resolution of the identity (RI),\citep{Vahtras1993_CPL_514}
is a pivotal technique in several recently developed computational
approaches for electronic structure theory. In the DF/RI approach,
two-electron integrals are expressed in terms of an auxiliary basis
set in Mulliken notation as\citep{Vahtras1993_CPL_514}
\begin{equation}
(\mu\nu|\sigma\rho)\approx\sum_{AB}(\mu\nu|A)(A|B)^{-1}(B|\sigma\rho),\label{eq:RI}
\end{equation}
where $\mu$, $\nu$, $\sigma$, and $\rho$ are orbital basis functions,
$A$ and $B$ are auxiliary basis functions, and $(A|B)^{-1}$ denotes
the $AB$ element of the inverse of the Coulomb overlap matrix. The
key of the DF/RI approach is that the set of atomic orbital (AO) products
$\mu\nu$ is heavily linearly dependent, and can be accurately approximated
with only a linearly scaling number of auxiliary functions $A$. 

Another key feature behind the usefulness of the DF/RI approach is
that the two-electron integrals in \eqref{RI} factorize. This factorization
can be exploited in many applications, as it allows the optimization
of the order of various contractions. For instance, the Coulomb matrix
\begin{equation}
J_{\mu\nu}=\sum_{\rho\sigma}(\mu\nu|\sigma\rho)P_{\sigma\rho},\label{eq:Jmat}
\end{equation}
where $\boldsymbol{P}$ denotes the density matrix, can be efficiently
evaluated with RI---yielding the RI-J scheme---in three steps:\citep{Eichkorn1995_CPL_283}
determination of the bare expansion coefficients $\gamma_{B}=\sum_{\rho\sigma}(B|\rho\sigma)P_{\sigma\rho}$,
projection into the orthonormal basis $\tilde{\gamma}_{A}=\sum_{B}(A|B)^{-1}\gamma_{B}$,
and assembly of the Coulomb matrix $J_{\mu\nu}=\sum_{A}(\mu\nu|A)\tilde{\gamma}_{A}$.
Exchange matrices $\boldsymbol{K}_{\sigma}$ are also often formed
with RI; the resulting RI-K algorithm is somewhat more complicated
than RI-J but efficient implementations have been developed also for
this case.\citep{Weigend2002_PCCP_4285,Manzer2015_JCP_24113} 

More generally, the transformation from the AO basis to the molecular
orbital (MO) basis that arises in various post-Hartree--Fock theories
\begin{equation}
(pq|rs)=\sum_{\mu\nu\sigma\rho}C_{\mu p}C_{\nu q}C_{\sigma r}C_{\rho s}(\mu\nu|\sigma\rho),\label{eq:mo-trans}
\end{equation}
where $\mathbf{C}$ is the matrix of MO coefficients, scales as $\mathcal{O}(N^{5})$
with the exact integrals $(\mu\nu|\sigma\rho)$ while the DF variant
only scales as $\mathcal{O}(N^{3})$, with $N$ denoting the number
of AO basis functions. DF thus yields significant speedups for calculations
employing second-order Møller--Plesset perturbation theory (MP2),\citep{Feyereisen1993_CPL_359,Weigend1998_CPL_143}
for instance. DF/RI is especially useful when the orbital basis set
is large, because the calculation of the exact two-electron integrals
as well as the integral transforms become slow in such cases.

Although most Gaussian-basis programs that support DF also also support
the use of exact two-electron integrals, there are also programs that
do not support exact integrals. For instance, the \Bagel{} program\citep{Shiozaki2018_WIRCMS_1331}
relies exclusively on DF in all calculations, as the DF integrals
can be kept in memory on modern hardware, and as the speedups DF offers
often have negligible effect on the accuracy of the calculation.\citep{Shiozaki2018_WIRCMS_1331}
For similar reasons, DF is the default mode of operation in the Gaussian-basis
\PsiFour{} program,\citep{Smith2020_JCP_184108} as well, even though
\PsiFour{} also supports the use of exact two-electron integrals.

Another important application of DF techniques can be found in quantum
chemistry programs employing atomic orbital basis sets other than
Gaussian-type orbitals, as the molecular two-electron integrals are
famously difficult to compute in such cases. Molecular calculations
with exact exchange as well as calculations at post-Hartree--Fock
levels of theory are made tractable by DF in combination with any
type of atomic orbital basis set, because the Coulomb potential of
the auxiliary functions is easy to evaluate numerically,\citep{Lehtola2019_IJQC_25968}
leaving only a three-dimensional integral which can be evaluated by
quadrature with Becke's multicenter method,\citep{Becke1988_JCP_2547}
for instance. Examples of such an approach are the Slater-type orbital
ADF program,\citep{Velde2001_JCC_931} as well as the numerical atomic
orbital \FHIaims{} program.\citep{Ren2012_NJP_53020} 

A central aspect of RI is the need for an auxiliary basis set. Traditionally,
an auxiliary basis set is optimized for each orbital basis set with
painstaking electronic structure calculations on a set of sample systems.\citep{Hill2013_IJQC_21}
For RI-J and RI-K, universal auxiliary basis sets by Weigend\citep{Weigend2006_PCCP_65,Weigend2008_JCC_167}
are typically used, even though the sets are formally tailored only
for the Karlsruhe def2 basis sets\citep{Weigend2005_PCCP_305} that
reach up to polarized quadruple-$\zeta$ quality. Post-Hartree--Fock
calculations in turn typically employ orbital-basis-specific auxiliary
basis sets parametrized by various authors.\citep{Weigend1998_CPL_143,Weigend2002_JCP_3175,Haettig2005_PCCP_59,Tanaka2013_JCC_2568,Hellweg2015_PCCP_1010}
Similarly, the Slater-type orbital basis sets used in ADF employ tailored
fitting basis sets.\citep{VanLenthe2003_JCC_56,Chong2004_JCC_6}

Although standard orbital and auxiliary basis sets are immensely useful
for several kinds of applications, they are not cost-efficient for
the reproduction of properties that are sensitive to aspects of the
wave function which do not have a significant impact on the total
energy. For instance, an accurate modeling of nuclear magnetic properties
requires specialized Gaussian basis sets with extremely tight exponents,\citep{Manninen2006_JCC_434,Jensen2006_JCTC_1360}
auxiliary basis sets for which may not be readily available. Another
example can be found in electron momentum densities which are sensitive
to the diffuse part of the wave function.\citep{Lehtola2011_PCCP_5630}
While specialized basis sets have likewise been developed for this
purpose,\citep{Lehtola2012_JCP_104105,Lehtola2013_JCP_44109} they
lack corresponding auxiliary basis sets. Yet another examples are
basis sets meant for core excitation spectroscopies,\citep{Ambroise2019_JCTC_325}
as well as the near-complete-basis-set-limit Gaussian basis sets for
the whole periodic table from H to Og formed from first principles,\citep{Lehtola2020_JCP_134108}
which similarly lack tailored auxiliary basis sets.

Automatic schemes for generating auxiliary basis sets from an orbital
basis set are thus immensely useful in many applications. \citet{Yang2007_JCP_74102}
proposed an algorithm for forming RI-J fitting sets from Gaussian
orbital basis sets by grouping together similar exponents arising
from all possible basis function products. \citet{Stoychev2017_JCTC_554}
proposed a heuristic algorithm that forms reasonably compact Gaussian
auxiliary basis sets that should be suitable for calculations both
at the self-consistent field\citep{Lehtola2020_M_1218} and post-Hartree--Fock
levels of theory. However, the schemes of \citeauthor{Yang2007_JCP_74102}
and \citeauthor{Stoychev2017_JCTC_554} have a number of adjustable
parameters and are therefore not completely black-box algorithms;
moreover, they are limited to the use of a Gaussian basis set.

Automatic algorithms are especially useful when they can be used with
any type of atomic basis set. \citet{Ren2012_NJP_53020} suggested
an approach where a Gram--Schmidt procedure is used to choose a linearly
independent set of basis function products as the auxiliary basis.
The algorithm of \citeauthor{Ren2012_NJP_53020} is in principle applicable
to any type of atomic basis set; their implementation in the \textsc{FHI-aims}
program\citep{Blum2009_CPC_2175} uses numerical atomic orbitals (see
\citeref{Lehtola2019_IJQC_25968} for discussion on various atomic
basis functions). However, the Gram--Schmidt procedure used by \citeauthor{Ren2012_NJP_53020}
is not guaranteed to process the auxiliary function candidates in
an optimal order, and thereby requires the use of an additional numerical
threshold for discarding functions that can be described sufficiently
well with the auxiliary radial functions previously included by the
procedure. Still, when applied to calculations with numerical atomic
orbital representations of standard Gaussian basis sets,\citep{Lehtola2019_IJQC_25968}
the procedure of \citeauthor{Ren2012_NJP_53020} has been found to
afford sub-meV-level RI errors in HF and MP2 total energies.\citep{Ren2012_NJP_53020,Ihrig2015_NJP_93020}

Aquilante et al.\citep{Aquilante2007_JCP_114107,Aquilante2009_JCP_154107}
pioneered an approach for forming auxiliary basis sets with a Cholesky
decomposition of the two-electron integral tensor.\citep{Beebe1977_IJQC_683}
Their algorithm is controlled by a single parameter: the Cholesky
decomposition threshold. \citet{Bostroem2009_JCTC_1545} showed that
the auxiliary sets from this approach can be made exact, that is,
the DF calculations can be made to agree with ones employing exact
two-electron integrals when the decomposition threshold is sufficiently
small. However, this approach that is implemented in the \textsc{OpenMolcas}
program\citep{Aquilante2020_JCP_214117} employs mixtures of Cartesian
and spherical basis functions as the auxiliary basis set even when
the orbital basis only has spherical functions,\bibnote{Roland Lindh, private communication, 2021}
thereby requiring a complicated approach which is not supported in
most quantum chemistry codes. Moreover, it turns out that unreliable
results are obtained if the mixed-form basis is used in either pure
Cartesian or pure spherical form, the latter being the standard choice
especially in the case of auxiliary basis sets. An automated algorithm
that produces spherical fitting functions would thus be highly useful.

In this work, we suggest a simplified approach for forming auxiliary
basis sets, which is a straightforward extension of the method developed
in \citerefs{Lehtola2019_JCP_241102} and \citenum{Lehtola2020_PRA_32504}
for solving issues with overcomplete orbital basis sets. Like the
scheme of \citet{Aquilante2009_JCP_154107}, the present algorithm
produces auxiliary basis sets that can be argued to be optimal; however,
unlike the scheme of \citeauthor{Aquilante2009_JCP_154107} the present
sets employ only spherical functions. At further variance to the scheme
of \citeauthor{Aquilante2009_JCP_154107}, which requires access to
the full two-electron integrals $(\mu\nu|\rho\sigma)$, the basic
version of the present approach only requires two-index integrals
$(A|B)$ like the scheme of \citet{Ren2012_NJP_53020} Our basic scheme
is obtained from the \citet{Ren2012_NJP_53020} scheme by replacing
the Gram--Schmidt method by the pivoted Cholesky method, which is
a numerically stable way to find an optimal set of auxiliary functions.\citep{Harbrecht2012_ANM_428,Lehtola2019_JCP_241102}
However, as we will show in this work, the present scheme can also
be combined with the Cholesky decomposition of the two-electron integrals
tensor $(\mu\nu|\rho\sigma)$, resulting in more cost-efficient auxiliary
basis sets.

The layout of the manuscript is the following. In \secref{Theory},
we discuss the full theory behind the present approach. We describe
the implementation of the method in \secref{Implementation}, and
describe the computational methods used in this work in \secref{compmet}.
The accuracy of the present method for RI-HF and RI-MP2 calculations
of total energies of the W4-17 test set of molecules\citep{Karton2017_JCC_2063}
is demonstrated in \secref{Results}; representative timings on commodity
hardware as well as further details of the algorithm are also discussed.
The article concludes in a summary and discussion in \secref{Summary}. 

\section{Theory \label{sec:Theory}}

Mulliken notation defines the electron repulsion integrals as
\begin{equation}
(\mu\nu|\rho\sigma)=\int\frac{\chi_{\mu}(\boldsymbol{r})\chi_{\nu}(\boldsymbol{r})\chi_{\rho}(\boldsymbol{r}')\chi_{\sigma}(\boldsymbol{r}')}{\left|\boldsymbol{r}-\boldsymbol{r}'\right|}{\rm d}^{3}r\,{\rm d}^{3}r',\label{eq:eri}
\end{equation}
where the basis functions have been assumed to be real. In this work,
both the orbital and the auxiliary basis functions are assumed to
be atomic
\begin{equation}
\chi_{\mu}(\boldsymbol{r})=R_{\mu}(r)Y_{l_{\mu}}^{m_{\mu}}(\hat{\boldsymbol{r}}),\label{eq:AO}
\end{equation}
where $R_{\mu}(r)$ is the radial function, and $Y_{l}^{m}$ are real-valued
spherical harmonics. In order for \eqref{RI} to be accurate, it is
seen from \eqref{eri} that the auxiliary basis set $\{A\}$ should
be able to represent all orbital basis function products $\{\chi_{\mu}\chi_{\nu}\}$
accurately; or, to be more precise, the potentials of $\{A\}$ should
be able to represent all the potentials of $\{\chi_{\mu}\chi_{\nu}\}$.
As the set $\{\chi_{\mu}\chi_{\nu}\}$ is heavily linearly dependent,
one should pick out the product functions in a way that spans all
possible degrees of freedom in the set as quickly as possible. This
is exactly what can be accomplished with a pivoted Cholesky decomposition.

The present approach to form auxiliary basis sets is thus the following.
The first step is to form all basis function products $\mu\nu$, yielding
the set of candidate auxiliary functions $\{\tilde{A}\}$. The products
are of the form $R_{\mu\nu}(r)Y_{L}^{M}(\hat{\boldsymbol{r}})$, where
$R_{\mu\nu}(r)=R_{\mu}(r)R_{\nu}(r)$ is a product of the radial basis
functions, while the angular part may be coupled to 
\begin{equation}
\left|l_{\mu}-l_{\nu}\right|\leq L\leq l_{\mu}-l_{\nu}\label{eq:am-coupling}
\end{equation}
 and $M=m_{\mu}+m_{\nu}$. Note that in this work, spherical auxiliary
functions are always used, as is the standard approach.\citep{Eichkorn1995_CPL_283}
Moreover, the present algorithm is shell-driven, that is, the auxiliary
functions are always handled one complete shell of functions ($M=-L,\dots,L$)
at a time. 

In the second step, the candidate functions' Coulomb overlap matrix
\begin{equation}
S_{\tilde{A}\tilde{B}}=(\tilde{A}|\tilde{B}).\label{eq:coulovl}
\end{equation}
is formed. The key point here is that the matrix $S_{\tilde{A}\tilde{B}}$
is block-diagonal in the angular momentum (see Appendix I): the dependence
on the magnetic quantum number $M$ can be omitted (justifying the
shell-based approach), leaving only dependence on the azimuthal quantum
number $L$. We can then form the matrix $S_{\tilde{A}\tilde{B}}$
separately for each angular momentum $L$ of the auxiliary basis set,
and the matrix elements only depend on the radial functions $\tilde{A}$
and $\tilde{B}$. As shown in Appendix I, the matrix elements in \eqref{coulovl}
have a trivial analytic form for Gaussian and Slater type orbitals,
whereas the integrals can be evaluated by quadrature when other kinds
of atomic basis functions such as numerical atomic orbitals are used.\citep{Lehtola2019_IJQC_25945,Lehtola2019_IJQC_25968}

Third, the method of \citerefs{Lehtola2019_JCP_241102} and \citenum{Lehtola2020_PRA_32504}
is used to pick a set of auxiliary functions $A$ from the pool $\tilde{A}$.
Since all basis function products $\mu\nu$ are included in the pool
$\tilde{A}$, and the procedure of ensures that all $\tilde{A}$ are
expressable in the basis $A$, since the residual norm of $\boldsymbol{S}$
is small after the Cholesky decomposition,\citep{Lehtola2019_JCP_241102}
the procedure yields an accurate auxiliary basis that is optimal.\citep{Harbrecht2012_ANM_428,Lehtola2019_JCP_241102}
The mathematical reasoning for the optimality is the following. The
Cholesky decomposition is given by\citep{Higham2009_WCS_251}
\begin{equation}
S_{\mu\nu}\approx\sum_{P}L_{\mu P}L_{\nu P}.\label{eq:cholesky}
\end{equation}
The pivoted algorithm, which is discussed extensively in \citeref{Harbrecht2012_ANM_428},
proceeds by iteratively adding new columns in $\boldsymbol{L}$ in
an optimal order. The pivot---the function to add---is the function
$A$ with the largest diagonal remainder, $(\boldsymbol{S}-\boldsymbol{L}\boldsymbol{L}^{\text{T}})_{AA}$.
This means that at every iteration, the pivoting procedure adds the
candidate function that is worst described by the set of the previous
pivot functions, until the predefined convergence threshold is reached.
Exactly this feature of the pivoted Cholesky algorithm is the reason
for its quick convergence,\citep{Harbrecht2012_ANM_428} affording
the optimal molecular basis sets pursued in \citerefs{Lehtola2019_JCP_241102}
and \citenum{Lehtola2020_PRA_32504}, as well as the optimal auxiliary
basis sets pursued in this work.

Because auxiliary functions are normalized, $(\tilde{A}|\tilde{A})=1$,
$\boldsymbol{S}$ has a unit diagonal as in the case of the usual
overlap matrix discussed in \citeref{Lehtola2019_JCP_241102}. Because
of the unit diagonal, the pivoted Cholesky algorithm starts out with
no information on the relative importance of the candidate basis functions.
The procedure of \citeref{Lehtola2019_JCP_241102} solved this issue
by presorting the basis functions by spatial extent, so that the functions
with the smallest spatial extent---which typically do not cause problems
with linear dependencies---are processed first. After the first iterations,
the diagonal remainders no longer equal unity, thus unleashing the
full power of the pivoted Cholesky algorithm.

However, ordering the functions by spatial extent is not the most
elegant solution in applications to the calculation of strongly repulsive
internuclear potentials, for example. In this application, the internuclear
distance can be small enough to cause overcompleteness even for the
tightest basis functions that represent core orbitals. A better solution
that requires only the overlap matrix $\boldsymbol{S}$ can be motivated
by the Gershgorin circle theorem.\citep{Gerschgorin1931_BldSlCdSmN_749} 

Per the note in \citeref{Lehtola2020_PRA_32504}, we sort the rows
and columns of $\boldsymbol{S}$ in increasing off-diagonal norm $S_{\tilde{A}}=\sum_{B}|S_{\tilde{A}\tilde{B}}|$.
This guarantees that the functions are fed into the pivoted Cholesky
procedure in an increasing order of potential linear dependencies.
The pivoted Cholesky is then initialized with the most independent
basis functions, which should anyway be included in the basis. This
should result in an optimal order in the pivoted Cholesky decomposition,
and result in the smallest possible subsets of basis functions chosen
by the pivoted Cholesky algorithm. This procedure also makes the pivoted
Cholesky algorithm self-sufficient, as in contrast of the original
approach used in \citerefs{Lehtola2019_JCP_241102} and \citenum{Lehtola2020_PRA_32504},
all the necessary information can be extracted from the decomposed
matrix, itself. In molecular applications along the lines of \citerefs{Lehtola2019_JCP_241102}
and \citenum{Lehtola2020_PRA_32504}, this ordering also accounts
for geometric effects: the basis functions that have the least overlap
with functions on other centers are processed first.

In the basic formalism, the present scheme only requires the computation
of the two-index integrals in \eqref{coulovl}. Alternatively, the
present method can also be combined with the Cholesky decomposition
of the electron repulsion integrals, in the lines of Aquilante et
al.\citep{Aquilante2007_JCP_114107,Aquilante2009_JCP_154107} Instead
of forming the pool of candidate auxiliary functions $\{\tilde{A}\}$
by generating \emph{all} products of basis functions in step 1, one
can instead include only those shell products that are chosen by the
pivoted Cholesky decomposition of the full atomic two-electron integrals
tensor\citep{Beebe1977_IJQC_683,Koch2003_JCP_9481}
\begin{equation}
(\mu\nu|\rho\sigma)\approx\sum_{P}L_{\mu\nu}^{P}L_{\rho\sigma}^{P};\label{eq:eri-cholesky}
\end{equation}
note the similarity of \eqref{RI, eri-cholesky} which is the motivation
for the approach of Aquilante et al\citep{Aquilante2007_JCP_114107,Aquilante2009_JCP_154107}
as well as the present approach. For each pivot index $\mu\nu$ chosen
by the decomposition, the product of the shells that the orbital basis
functions $\mu$ and $\nu$ belong to is added to the list of candidate
shells. 

Using the Cholesky decomposition of the two-electron integrals to
choose the pool of candidate functions in step 1 leads to smaller
auxiliary basis sets than using all possible atomic-orbital products
as candidates, as will be shown later in this work. The difference
is especially noticeable for high-angular-momentum orbital basis sets.
For this reason, we will call these auxiliary basis sets \emph{reduced
auxiliary basis sets}, whereas the auxiliary basis sets formed from
the consideration of all possible product functions are termed \emph{full
auxiliary basis sets}.

Although, in principle, different thresholds could be used for the
pivoted Cholesky decomposition of the two-electron integral tensor
and the pivoted Cholesky decomposition of the candidate auxiliary
basis functions' Coulomb overlap matrix (\eqref{coulovl}) which is
used to choose the auxiliary basis functions in this work, for simplicity
we have opted to use the same threshold $\tau$ for both, as thresholds
of $\tau=10^{-3}$ to $\tau=10^{-7}$ are reasonable for either part
of the problem. Decreasing the threshold $\tau$ then simultaneously
results in more candidate functions from the decomposition of the
two-electron integrals tensor, as well as a tighter criterion for
the representation of all of the candidate functions by the chosen
auxiliary basis set.

\section{Implementation \label{sec:Implementation}}

Both the full and the reduced variants of the present approach were
implemented for Gaussian basis sets in ERKALE\citep{Lehtola2012_JCC_1572,Lehtola2018__a}
by building on top of the existing implementation\citep{Lehtola2016_JCTC_3195}
of the Cholesky decomposition of two-electron integrals for molecular
calculations.\citep{Beebe1977_IJQC_683,Koch2003_JCP_9481} For clarity,
the present scheme for forming auxiliary basis sets discussed in \secref{Theory}
is summarized in \algref{Flowchart-algorithm.}; details of the algorithm
that are specific to the present use of Gaussian basis sets will be
discussed below. For comparison, the AutoAbs and AutoAux methods of
\citet{Yang2007_JCP_74102} and \citet{Stoychev2017_JCTC_554}, respectively,
were implemented in the Python backend library of the Basis Set Exchange\citep{Pritchard2019_JCIM_4814}
as part of this work. 

The use of Gaussian basis sets implies some limitations in the present
proof-of-concept study, which need to be documented. First, we will
only consider fully uncontracted orbital basis sets in the construction
of the auxiliary basis, as this makes the algorithm simpler; this
means that the orbital basis set is always decontracted before building
the auxiliary basis set. (Note that applications to numerical atomic
orbitals will not require such a step, because the product of radial
functions $R_{\mu\nu}(r)=R_{\mu}(r)R_{\nu}(r)$ is simply a new numerical
atomic orbital, although potential issues with aliasing\citep{Lehtola2019_IJQC_25968}
should be kept in mind.)

Second, Gaussian-type orbitals have a fixed radial form
\begin{equation}
R_{nl}^{\text{GTO}}(r)=r^{l}\exp(-\alpha_{nl}r^{2})\label{eq:gto}
\end{equation}
which needs to be considered in the assembly of the pool of candidate
functions. The product of two Gaussian radial functions
\begin{equation}
R_{\mu}(r)R_{\nu}(r)=r^{l_{\mu}+l_{\nu}}e^{-(\alpha_{\mu}+\alpha_{\nu})r^{2}}\label{eq:gto-prod}
\end{equation}
spawns candidates to multiple angular momentum channels $L$ according
to \eqref{am-coupling}; each candidate has the form of \eqref{gto}.
For example, an F function ($l=3$) with exponent $\alpha$ can couple
with itself to $0\leq L\leq6$. While the real product radial function
is $r^{6}\exp(-2\alpha r^{2})$, the form of the actual auxiliary
functions $r^{L}\exp(-2\alpha r^{2})$ depends on the value of the
coupled angular momentum $L$. The real product function is more diffuse
than the candidate for $L\leq\max L$; this difference has to be captured
by the algorithm. Following \citet{Stoychev2017_JCTC_554}, we solve
this issue by employing an effective exponent $\alpha{}_{L}$ for
the candidate function $r^{L}\exp(-2\alpha{}_{L}r^{2})$ so that the
expectation value of $\langle r\rangle$ for the generated auxiliary
function candidate agrees with the expectation value of $\langle r\rangle$
computed with the true radial product function; this procedure is
documented in Appendix II.

\begin{algorithm}
\begin{enumerate}
\item Obtain the orbital basis set defined by the radial functions $R_{nl}(r)$
with angular momentum $l$, $\chi_{\alpha}=R_{n_{\alpha}l_{\alpha}}(r)Y_{l_{\alpha}}^{m_{\alpha}}(\boldsymbol{r})$.
Each radial function $R_{nl}(r)$ thus corresponds to the $2l+1$
functions with $m=-l,\dots,l$.\label{enu:Obtain-the-orbital}
\item Form all products of orbital basis functions $\chi_{\alpha}\chi_{\beta}$;
this yields the full set of candidate functions. The reduced set is
obtained by considering only those products of basis functions that
appear as pivot indices in the Cholesky decomposition of the two-electron
integral tensor, instead.\\
\medskip{}
The radial part of the basis function products is $R_{n_{\alpha}l_{\alpha}}(r)R_{n_{\beta}l_{\beta}}(r)$,
and the angular part is $Y_{l_{\alpha}}^{m_{\alpha}}(\boldsymbol{r})Y_{l_{\beta}}^{m_{\beta}}(\boldsymbol{r})$.
As the product of spherical harmonics is closed, the angular part
can be rewritten in terms of spherical harmonics in the range $|l_{\alpha}-l_{\beta}|\leq L\leq l_{\alpha}+l_{\beta}$.
The candidate auxiliary functions are then identified as $R_{n_{\alpha}l_{\alpha}}(r)R_{n_{\beta}l_{\beta}}(r)Y_{L}^{M}$.\\
\medskip{}
 If basis functions with a rigid analytic form are employed, like
the even-tempered Gaussian functions of this work, the radial function
products $R_{n_{\alpha}l_{\alpha}}(r)R_{n_{\beta}l_{\beta}}(r)$ are
replaced everywhere in the algorithm by an effective radial function
$R_{n_{\alpha\beta}L}(r)$ for each value of $L$, as discussed in
Appendix II.\label{enu:Form-all-products}
\item As Appendix I demonstrates, the Coulomb overlap matrix of the candidate
functions is diagonal in $L$, meaning that all the candidate radial
functions $R_{n_{\alpha}l_{\alpha}}(r)R_{n_{\beta}l_{\beta}}(r)$
corresponding to a given final value of $L$ can be considered together.
Therefore, the auxiliary basis can be determined separately for each
value of $L$:
\begin{enumerate}
\item Form the Coulomb overlap matrix $S_{\tilde{A}\tilde{B}}=(\tilde{A}|\tilde{B})$
for the candidate functions corresponding to the given $L$ shell.
Because the set of candidate functions is heavily overcomplete, this
matrix is extremely ill-conditioned.\label{enu:Form-the-Coulomb}
\item Use the pivoted Cholesky decomposition of \citerefs{Lehtola2019_JCP_241102}
and \citenum{Lehtola2020_PRA_32504} to choose a linearly independent
set of auxiliary functions from the pool of candidate functions with
the given $L$.\label{enu:Use-the-pivoted}
\begin{enumerate}
\item Order the candidate functions in increasing values of off-diagonal
overlap, $s_{\tilde{A}}=\sum_{\tilde{B}\neq\tilde{A}}|S_{\tilde{A}\tilde{B}}|$.
This guarantees that the most linearly independent candidate functions
are picked first.\citep{Lehtola2020_PRA_32504} \label{enu:Since-the-auxiliary}
\item Perform the pivoted Cholesky decomposition\citep{Harbrecht2012_ANM_428}
for the reordered Coulomb overlap matrix $\boldsymbol{S}$ up to the
specified decomposition threshold $\tau$ with the LAPACK {[}Linear
Algebra PACKage{]} routine \texttt{dpstrf()}, for example, and store
the returned pivot indices. \label{enu:A-pivoted-Cholesky}
\item Map the pivot indices back to the original indexing of the candidate
functions.
\end{enumerate}
\item Add the candidate radial functions corresponding to the used pivot
indices to the auxiliary basis.\label{enu:The-pivot-indices}
\end{enumerate}
\end{enumerate}
\caption{Summary of the algorithm for forming the auxiliary basis starting
from the orbital basis specified for an atom. \label{alg:Flowchart-algorithm.}}
\end{algorithm}

\section{Computational Methods \label{sec:compmet}}

As was already mentioned in the Introduction, the W4-17 database of
molecules\citep{Karton2017_JCC_2063} is used to test the auxiliary
basis sets. RI-HF and RI-MP2 calculations were performed with the
double-$\zeta$ (2ZaPa-NR) to quintuple-$\zeta$ (5ZaPa-NR) orbital
basis sets of \citet{Ranasinghe2013_JCP_144104} using \textsc{Psi4}.\citep{Smith2020_JCP_184108}
The orbital basis sets and the corresponding AutoAux sets were obtained
from the Basis Set Exchange.\citep{Pritchard2019_JCIM_4814} Conventional
HF and MP2 calculations were carried out with \textsc{Gaussian'09};\citep{Frisch2009__a}
manipulations to the input basis were disabled with the setting \texttt{IOp(3/60=-1)}.
A basis set linear dependence threshold of $10^{-7}$ was used in
both programs; program defaults were used otherwise. To ensure the
validity of RI-MP2, the multireference part of the database is excluded
from the analysis, as is the hydrogen atom that has no correlation
energy in RI-MP2 and for which the exchange and Coulomb terms cancel
out in RI-HF. Core orbitals were frozen in all MP2 calculations.

\section{Results \label{sec:Results}}

\subsection{Full auxiliary basis versus reduced auxiliary basis \label{subsec:Full-auxiliary-basis}}

As was mentioned in \secref{Theory}, the present approach can be
implemented in two ways. The first is to pick the auxiliary basis
set from the set of all possible orbital products (full approach);
the second is to pick the auxiliary basis set from only the set of
orbital products that are picked up by the Cholesky decomposition
of the two-electron integral tensor (reduced approach). To illustrate
the difference between these two approaches, the compositions of the
auxiliary basis sets arising from the 2ZaPa-NR and 5ZaPa-NR orbital
basis sets are shown in \tabref{2zapa, 5zapa}, respectively.

The data in \tabref{2zapa} suggest that the differences between the
full and the reduced auxiliary basis sets are small for small orbital
basis sets. In contrast, significant differences are observed in the
case of large orbital basis sets exemplified by \tabref{5zapa}. 

Even though the compositions of the full and reduced auxiliary basis
sets are similar at both small and large angular momentum, the differences
at intermediate angular momentum are huge. For instance, the reduced
auxiliary basis contains roughly just one half the number of f, g,
and h functions of the full auxiliary basis. This difference between
the two approaches is easy to understand. Products of high-angular
momentum basis functions can couple to many values of angular momentum.
For instance, the product of two F functions ($l=3$) may have components
ranging from S functions to I functions ($l=6$). If the coupling
coefficients are small at high angular momenta, the value of the two-electron
integral may be dominated by contributions from the low-angular-momentum
functions. The decomposition of the two-electron integrals tensor
picks out only those basis function products that are necessary to
describe the two-electron integrals, leading to the observed improvement
in efficiency.

\begin{table*}
\begin{tabular}{lllll}
{\footnotesize{}atom} & {\footnotesize{}primitive orbital basis} & {\footnotesize{}AutoAux basis} & {\footnotesize{}full auxiliary basis} & {\footnotesize{}reduced auxiliary basis}\tabularnewline
\hline 
\hline 
{\footnotesize{}H} & {\footnotesize{}6s1p} & {\footnotesize{}12s2p2d} & {\footnotesize{}12s6p1d} & {\footnotesize{}10s6p1d}\tabularnewline
{\footnotesize{}He} & {\footnotesize{}7s1p} & {\footnotesize{}10s2p2d} & {\footnotesize{}13s6p1d} & {\footnotesize{}11s6p1d}\tabularnewline
{\footnotesize{}Li} & {\footnotesize{}9s5p1d} & {\footnotesize{}16s13p12d2f} & {\footnotesize{}22s21p16d5f1g} & {\footnotesize{}20s16p13d5f1g}\tabularnewline
{\footnotesize{}Be} & {\footnotesize{}10s5p1d} & {\footnotesize{}16s13p12d2f} & {\footnotesize{}22s21p17d5f1g} & {\footnotesize{}21s17p14d5f1g}\tabularnewline
{\footnotesize{}B} & {\footnotesize{}10s6p1d} & {\footnotesize{}16s13p12d2f} & {\footnotesize{}22s21p18d6f1g} & {\footnotesize{}21s19p15d6f1g}\tabularnewline
{\footnotesize{}C} & {\footnotesize{}11s6p1d} & {\footnotesize{}16s13p12d2f} & {\footnotesize{}23s23p19d6f1g} & {\footnotesize{}22s19p15d6f1g}\tabularnewline
{\footnotesize{}N} & {\footnotesize{}11s7p1d} & {\footnotesize{}16s13p12d2f} & {\footnotesize{}24s24p21d7f1g} & {\footnotesize{}23s21p16d6f1g}\tabularnewline
{\footnotesize{}O} & {\footnotesize{}11s7p1d} & {\footnotesize{}16s13p12d2f} & {\footnotesize{}24s23p21d7f1g} & {\footnotesize{}23s22p16d7f1g}\tabularnewline
{\footnotesize{}F} & {\footnotesize{}11s8p1d} & {\footnotesize{}16s13p12d2f} & {\footnotesize{}24s24p22d8f1g} & {\footnotesize{}23s22p18d7f1g}\tabularnewline
{\footnotesize{}Ne} & {\footnotesize{}11s8p1d} & {\footnotesize{}16s13p13d2f} & {\footnotesize{}24s24p22d8f1g} & {\footnotesize{}23s22p18d7f1g}\tabularnewline
{\footnotesize{}Na} & {\footnotesize{}14s9p2d} & {\footnotesize{}21s19p18d5f} & {\footnotesize{}31s31p28d16f3g} & {\footnotesize{}29s28p24d12f3g}\tabularnewline
{\footnotesize{}Mg} & {\footnotesize{}15s9p2d} & {\footnotesize{}21s18p16d5f} & {\footnotesize{}31s31p26d16f3g} & {\footnotesize{}28s27p22d12f3g}\tabularnewline
{\footnotesize{}Al} & {\footnotesize{}14s10p2d} & {\footnotesize{}20s17p16d5f} & {\footnotesize{}30s30p27d17f3g} & {\footnotesize{}30s28p23d12f3g}\tabularnewline
{\footnotesize{}Si} & {\footnotesize{}14s10p2d} & {\footnotesize{}20s17p16d5f} & {\footnotesize{}29s30p27d17f3g} & {\footnotesize{}28s28p24d12f3g}\tabularnewline
{\footnotesize{}P} & {\footnotesize{}14s10p2d} & {\footnotesize{}19s16p15d5f} & {\footnotesize{}29s29p27d17f3g} & {\footnotesize{}28s27p23d12f3g}\tabularnewline
{\footnotesize{}S} & {\footnotesize{}14s10p2d} & {\footnotesize{}19s16p15d5f} & {\footnotesize{}29s29p27d17f3g} & {\footnotesize{}28s26p23d13f3g}\tabularnewline
{\footnotesize{}Cl} & {\footnotesize{}14s10p2d} & {\footnotesize{}19s16p15d5f} & {\footnotesize{}29s29p27d17f3g} & {\footnotesize{}27s26p24d13f3g}\tabularnewline
{\footnotesize{}Ar} & {\footnotesize{}14s10p2d} & {\footnotesize{}19s16p15d5f} & {\footnotesize{}29s29p27d16f3g} & {\footnotesize{}28s26p23d12f3g}\tabularnewline
\end{tabular}

\caption{Composition of 2ZaPa-NR and the resulting AutoAux basis as well as
the full and reduced auxiliary basis sets with threshold $\tau=10^{-7}$.\label{tab:2zapa}}
\end{table*}

\begin{sidewaystable*}
\begin{tabular}{lllll}
{\footnotesize{}atom} & {\footnotesize{}primitive orbital basis} & {\footnotesize{}AutoAux basis} & {\footnotesize{}full auxiliary basis} & {\footnotesize{}reduced auxiliary basis}\tabularnewline
\hline 
\hline 
{\footnotesize{}H} & {\footnotesize{}15s5p4d3f1g} & {\footnotesize{}17s8p7d6f5g5h} & {\footnotesize{}22s19p19d18f15g11h8i4j1k} & {\footnotesize{}20s17p15d15f12g10h8i4j1k}\tabularnewline
{\footnotesize{}He} & {\footnotesize{}16s5p4d3f1g} & {\footnotesize{}16s8p7d6f6g5h} & {\footnotesize{}24s22p22d23f18g11h9i4j1k} & {\footnotesize{}23s17p17d14f13g11h9i4j1k}\tabularnewline
{\footnotesize{}Li} & {\footnotesize{}18s12p5d4f3g1h} & {\footnotesize{}16s14p13d7f6g5h4i} & {\footnotesize{}30s29p25d25f25g23h14i9j7k4l1m} & {\footnotesize{}28s24p20d16f14g12h10i9j7k4l1m}\tabularnewline
{\footnotesize{}Be} & {\footnotesize{}19s12p5d4f3g1h} & {\footnotesize{}16s13p12d7f6g6h3i} & {\footnotesize{}32s29p26d25f26g24h13i10j8k4l1m} & {\footnotesize{}29s25p20d17f15g13h11i10j7k4l1m}\tabularnewline
{\footnotesize{}B} & {\footnotesize{}19s13p5d4f3g1h} & {\footnotesize{}16s14p13d7f7g6h4i} & {\footnotesize{}32s31p27d28f27g26h14i10j8k4l1m} & {\footnotesize{}30s26p22d19f17g14h11i10j8k4l1m}\tabularnewline
{\footnotesize{}C} & {\footnotesize{}20s14p5d4f3g1h} & {\footnotesize{}16s13p13d7f7g6h4i} & {\footnotesize{}31s32p28d28f29g27h15i10j8k4l1m} & {\footnotesize{}31s28p23d18f16g14h11i10j8k4l1m}\tabularnewline
{\footnotesize{}N} & {\footnotesize{}20s16p5d4f3g1h} & {\footnotesize{}17s15p14d7f7g6h4i} & {\footnotesize{}32s32p29d29f30g28h17i11j8k4l1m} & {\footnotesize{}32s29p26d21f17g16h13i11j8k4l1m}\tabularnewline
{\footnotesize{}O} & {\footnotesize{}20s15p5d4f3g1h} & {\footnotesize{}16s14p13d7f7g6h4i} & {\footnotesize{}32s31p29d29f30g28h17i11j8k4l1m} & {\footnotesize{}32s28p25d21f17g16h14i11j8k4l1m}\tabularnewline
{\footnotesize{}F} & {\footnotesize{}20s16p5d4f3g1h} & {\footnotesize{}16s14p13d7f7g6h4i} & {\footnotesize{}33s32p30d29f30g28h17i11j8k4l1m} & {\footnotesize{}32s31p26d22f18g16h14i11j8k4l1m}\tabularnewline
{\footnotesize{}Ne} & {\footnotesize{}20s17p5d4f3g1h} & {\footnotesize{}17s14p14d7f7g6h4i} & {\footnotesize{}32s32p31d30f29g30h17i11j8k4l1m} & {\footnotesize{}31s30p28d22f18g16h14i11j8k4l1m}\tabularnewline
{\footnotesize{}Na} & {\footnotesize{}23s18p6d4f3g1h} & {\footnotesize{}21s18p17d10f10g9h5i} & {\footnotesize{}36s38p36d35f36g35h20i10j7k4l1m} & {\footnotesize{}37s35p30d21f16g14h12i10j6k4l1m}\tabularnewline
{\footnotesize{}Mg} & {\footnotesize{}23s18p6d4f3g1h} & {\footnotesize{}20s18p17d10f9g9h4i} & {\footnotesize{}35s37p35d35f35g35h20i10j7k4l1m} & {\footnotesize{}36s35p30d20f14g13h12i10j7k4l1m}\tabularnewline
{\footnotesize{}Al} & {\footnotesize{}23s19p6d4f3g1h} & {\footnotesize{}20s17p16d10f9g9h4i} & {\footnotesize{}35s38p36d37f36g35h19i10j8k4l1m} & {\footnotesize{}35s35p31d20f16g13h11i10j8k4l1m}\tabularnewline
{\footnotesize{}Si} & {\footnotesize{}23s19p6d4f3g1h} & {\footnotesize{}20s17p16d10f9g9h5i} & {\footnotesize{}35s38p36d36f36g34h20i11j8k4l1m} & {\footnotesize{}35s35p30d21f15g14h13i11j8k4l1m}\tabularnewline
{\footnotesize{}P} & {\footnotesize{}23s19p6d4f3g1h} & {\footnotesize{}19s17p16d10f9g8h5i} & {\footnotesize{}35s36p36d36f35g35h20i11j7k4l1m} & {\footnotesize{}34s35p31d22f16g14h13i11j7k4l1m}\tabularnewline
{\footnotesize{}S} & {\footnotesize{}23s19p6d4f3g1h} & {\footnotesize{}19s16p16d10f9g8h5i} & {\footnotesize{}34s37p36d36f36g35h21i11j8k4l1m} & {\footnotesize{}35s36p30d21f16g14h13i11j8k4l1m}\tabularnewline
{\footnotesize{}Cl} & {\footnotesize{}23s19p6d4f3g1h} & {\footnotesize{}19s16p15d9f9g8h5i} & {\footnotesize{}34s37p35d36f36g35h21i11j8k4l1m} & {\footnotesize{}35s34p30d21f16g15h13i11j8k4l1m}\tabularnewline
{\footnotesize{}Ar} & {\footnotesize{}23s19p6d4f3g1h} & {\footnotesize{}20s17p16d9f9g8h5i} & {\footnotesize{}35s37p36d36f35g34h21i11j8k4l1m} & {\footnotesize{}34s35p30d21f16g14h13i11j8k4l1m}\tabularnewline
\end{tabular}

\caption{Composition of 5ZaPa-NR and the resulting AutoAux basis as well as
the full and reduced auxiliary basis sets with threshold $\tau=10^{-7}$.\label{tab:5zapa}}
\end{sidewaystable*}

The accuracy of the full and reduced auxiliary basis sets can be quantified
by comparing the resulting errors in the diagonal of the electron
repulsion integral tensor $(\mu\nu|\mu\nu)$ for which the error is
negative definite: $(\mu\nu|\mu\nu)$ is essentially the self-interaction
energy of the electron density given by $\chi_{\mu}(\boldsymbol{r})\chi_{\nu}(\boldsymbol{r})$.
The diagonal integral evaluated with RI using \eqref{RI}, $\widetilde{(\mu\nu|\mu\nu)}$,
is smaller than the exact value; giving rise to an error metric
\begin{equation}
\Delta=\sum_{\mu\nu}\left[(\mu\nu|\mu\nu)-\widetilde{(\mu\nu|\mu\nu)}\right]\geq0\label{eq:Delta}
\end{equation}
which we have implemented in ERKALE.\citep{Lehtola2012_JCC_1572,Lehtola2018__a} 

\Eqref{Delta} can also be used to compare other types of auxiliary
basis sets in a first-principles fashion. Such a comparison is shown
in \tabref{diagonal-error} for the Fe def2-QZVP orbital basis set;\citep{Weigend2003_JCP_12753}
other choices for the orbital basis and other elements yield similar
results. The comparison includes the universal auxiliary basis sets
for RI-J\citep{Weigend2006_PCCP_65} and RI-JK\citep{Weigend2008_JCC_167}
calculations, the auxiliary basis set for MP2 calculations with the
slightly larger def2-QZVPP orbital basis set,\citep{Haettig2005_PCCP_59}
as well as automatically generated auxiliary basis sets for the def2-QZVP
orbital basis using the AutoAbs\citep{Yang2007_JCP_74102} and AutoAux\citep{Stoychev2017_JCTC_554}
approaches and the full and reduced algorithms of this work. The universal
RI-J and RI-JK auxiliary sets and the MP2 auxiliary set were fully
uncontracted for an unbiased comparison with the automatically generated
basis sets. 

The RI-J sets are aimed at reproducing the Coulomb potential, which
tends to be spherically symmetric; thus already low-angular-momentum
integrals like (SP|SP) have significant errors. Likewise, the AutoAbs
sets target Coulomb integrals. Even though some of the low-angular-momentum
integrals are reproduced more accurately by AutoAbs than by the universal
RI-J auxiliary sets, the AutoAbs method still exhibits significant
errors overall, only halving the total error observed for the universal
RI-J auxiliary set.

The RI-JK sets, in turn, have to be able to describe individual orbital
densities; this is reflected in smaller errors in small-angular-momentum
integrals that describe interactions of the occupied shells in the
iron atom. The RI-JK set still has large errors for high-angular-momentum
integrals.

Another interesting point of comparison is offered by the auxiliary
basis set for MP2 calculations with the def2-QZVPP orbital basis set
(def2-QZVPP-RIFIT). As def2-QZVPP is def2-QZVP with additional polarization
functions, the auxiliary basis is sensible also for the studied def2-QZVP
orbital basis set. Interesting differences between the accuracy of
the MP2 auxiliary basis set and the RI-JK sets can be observed. The
MP2 auxiliary basis is noticeably less accurate than the RI-JK set
for (SS|SS) and (PD|PD) integrals, which describe interactions between
occupied orbitals in the iron atom. But, the MP2 auxiliary basis is
noticeably more accurate than the RI-JK set for (DF|DF), (FF|FF),
(SG|SG), (PG|PG), and (DG|DG) integrals that arise from polarization
as well as electron correlation. The accuracies of the two auxiliary
basis sets for other classes of integrals are more or less similar.
The total error for the def2-QZVPP-RIFIT auxiliary basis is slightly
smaller than for the universal RI-JK set.

Moving onto the general-use basis sets generated by automatic approaches,
the AutoAux sets are tailored for computational efficiency at various
levels of theory, minimizing the number of auxiliary functions without
sacrificing the resulting accuracy; AutoAux especially truncates high-angular-momentum
basis functions which are typically not necessary for global fitting
approaches. Its total error is smaller than that of the def2-QZVPP-RIFIT
auxiliary basis, scoring in at half the error of the universal RI-JK
set, and the remaining error is dominated by deficiencies at large
angular momentum.

In contrast, the automatically generated auxiliary basis sets from
the present approach are able to describe all basis function products
that arise from the specified orbital basis, as demonstrated by the
small errors in the integrals. Even the (GG|GG) integrals that describe
the interactions of the second shell of polarization functions, which
are not expected to be important in HF or MP2 calculations, are captured
accurately by the present scheme, demonstrating its reliability for
any property and any level of theory.

Because the full and reduced schemes appear to afford similar levels
of accuracy despite the smaller number of functions involved in the
reduced scheme, we will only consider reduced auxiliary basis sets
in the remainder of this work unless explicitly specified otherwise. 

\begin{sidewaystable*}
\begin{centering}
\begin{tabular}{l|lllllll}
 & {\scriptsize{}def2-universal-jfit} & {\scriptsize{}def2-universal-jkfit} & {\scriptsize{}def2-QZVPP-RIFIT} & {\scriptsize{}AutoAbs} & {\scriptsize{}AutoAux} & {\scriptsize{}full $\tau=10^{-7}$} & {\scriptsize{}reduced $\tau=10^{-7}$}\tabularnewline
{\scriptsize{}integral type} & {\scriptsize{}19s5p5d3f3g} & {\scriptsize{}19s14p12d10f7g3h1i} & {\scriptsize{}15s13p11d8f7g5h2i} & {\scriptsize{}24s19p14d8f7g} & {\scriptsize{}22s19p18d17f16g6h4i} & {\scriptsize{}36s38p39d38f37g24h12i4j1k} & {\scriptsize{}38s36p33d28f25g15h10i4j1k}\tabularnewline
\hline 
{\scriptsize{}(SS|SS)} & {\scriptsize{}$6.878\times10^{-5}$} & {\scriptsize{}$1.283\times10^{-4}$} & {\scriptsize{}$1.942\times10^{-3}$} & {\scriptsize{}$7.274\times10^{-5}$} & {\scriptsize{}$9.411\times10^{-6}$} & {\scriptsize{}$1.217\times10^{-5}$} & {\scriptsize{}$9.466\times10^{-6}$}\tabularnewline
{\scriptsize{}(SP|SP)} & {\scriptsize{}$2.661\times10^{0}$} & {\scriptsize{}$2.864\times10^{-4}$} & {\scriptsize{}$2.833\times10^{-4}$} & {\scriptsize{}$2.400\times10^{-4}$} & {\scriptsize{}$3.692\times10^{-6}$} & {\scriptsize{}$1.233\times10^{-6}$} & {\scriptsize{}$6.744\times10^{-7}$}\tabularnewline
{\scriptsize{}(PP|PP)} & {\scriptsize{}$1.086\times10^{0}$} & {\scriptsize{}$8.635\times10^{-3}$} & {\scriptsize{}$1.098\times10^{-2}$} & {\scriptsize{}$2.707\times10^{-3}$} & {\scriptsize{}$1.075\times10^{-4}$} & {\scriptsize{}$5.518\times10^{-6}$} & {\scriptsize{}$9.908\times10^{-7}$}\tabularnewline
{\scriptsize{}(SD|SD)} & {\scriptsize{}$8.620\times10^{-2}$} & {\scriptsize{}$2.051\times10^{-4}$} & {\scriptsize{}$4.392\times10^{-4}$} & {\scriptsize{}$7.592\times10^{-4}$} & {\scriptsize{}$9.842\times10^{-6}$} & {\scriptsize{}$2.279\times10^{-7}$} & {\scriptsize{}$1.294\times10^{-7}$}\tabularnewline
{\scriptsize{}(PD|PD)} & {\scriptsize{}$5.723\times10^{-1}$} & {\scriptsize{}$1.221\times10^{-3}$} & {\scriptsize{}$4.604\times10^{-2}$} & {\scriptsize{}$6.614\times10^{-2}$} & {\scriptsize{}$1.207\times10^{-4}$} & {\scriptsize{}$2.826\times10^{-6}$} & {\scriptsize{}$1.438\times10^{-6}$}\tabularnewline
{\scriptsize{}(DD|DD)} & {\scriptsize{}$5.656\times10^{-1}$} & {\scriptsize{}$1.394\times10^{-1}$} & {\scriptsize{}$9.993\times10^{-2}$} & {\scriptsize{}$8.098\times10^{-2}$} & {\scriptsize{}$2.926\times10^{-4}$} & {\scriptsize{}$7.261\times10^{-5}$} & {\scriptsize{}$2.361\times10^{-5}$}\tabularnewline
{\scriptsize{}(SF|SF)} & {\scriptsize{}$5.295\times10^{-1}$} & {\scriptsize{}$1.265\times10^{-3}$} & {\scriptsize{}$9.474\times10^{-4}$} & {\scriptsize{}$1.105\times10^{-2}$} & {\scriptsize{}$4.507\times10^{-5}$} & {\scriptsize{}$5.630\times10^{-7}$} & {\scriptsize{}$2.322\times10^{-6}$}\tabularnewline
{\scriptsize{}(PF|PF)} & {\scriptsize{}$1.437\times10^{-1}$} & {\scriptsize{}$1.702\times10^{-3}$} & {\scriptsize{}$1.002\times10^{-3}$} & {\scriptsize{}$7.409\times10^{-2}$} & {\scriptsize{}$1.149\times10^{-4}$} & {\scriptsize{}$1.216\times10^{-6}$} & {\scriptsize{}$1.413\times10^{-6}$}\tabularnewline
{\scriptsize{}(DF|DF)} & {\scriptsize{}$2.490\times10^{0}$} & {\scriptsize{}$6.310\times10^{-2}$} & {\scriptsize{}$1.113\times10^{-2}$} & {\scriptsize{}$2.046\times10^{0}$} & {\scriptsize{}$1.059\times10^{-2}$} & {\scriptsize{}$4.809\times10^{-5}$} & {\scriptsize{}$4.257\times10^{-5}$}\tabularnewline
{\scriptsize{}(FF|FF)} & {\scriptsize{}$2.191\times10^{0}$} & {\scriptsize{}$9.606\times10^{-1}$} & {\scriptsize{}$3.803\times10^{-1}$} & {\scriptsize{}$2.018\times10^{0}$} & {\scriptsize{}$3.901\times10^{-2}$} & {\scriptsize{}$3.391\times10^{-4}$} & {\scriptsize{}$3.498\times10^{-4}$}\tabularnewline
{\scriptsize{}(SG|SG)} & {\scriptsize{}$1.986\times10^{-2}$} & {\scriptsize{}$6.502\times10^{-4}$} & {\scriptsize{}$1.344\times10^{-4}$} & {\scriptsize{}$9.036\times10^{-4}$} & {\scriptsize{}$3.083\times10^{-5}$} & {\scriptsize{}$3.469\times10^{-8}$} & {\scriptsize{}$8.586\times10^{-8}$}\tabularnewline
{\scriptsize{}(PG|PG)} & {\scriptsize{}$7.234\times10^{-1}$} & {\scriptsize{}$1.180\times10^{-2}$} & {\scriptsize{}$1.146\times10^{-3}$} & {\scriptsize{}$5.159\times10^{-1}$} & {\scriptsize{}$1.807\times10^{-3}$} & {\scriptsize{}$1.354\times10^{-6}$} & {\scriptsize{}$4.869\times10^{-7}$}\tabularnewline
{\scriptsize{}(DG|DG)} & {\scriptsize{}$7.276\times10^{-1}$} & {\scriptsize{}$1.948\times10^{-1}$} & {\scriptsize{}$5.872\times10^{-2}$} & {\scriptsize{}$6.714\times10^{-1}$} & {\scriptsize{}$1.391\times10^{-2}$} & {\scriptsize{}$4.954\times10^{-6}$} & {\scriptsize{}$3.183\times10^{-6}$}\tabularnewline
{\scriptsize{}(FG|FG)} & {\scriptsize{}$1.604\times10^{0}$} & {\scriptsize{}$6.855\times10^{-1}$} & {\scriptsize{}$6.610\times10^{-1}$} & {\scriptsize{}$1.267\times10^{0}$} & {\scriptsize{}$6.576\times10^{-1}$} & {\scriptsize{}$4.911\times10^{-5}$} & {\scriptsize{}$7.325\times10^{-5}$}\tabularnewline
{\scriptsize{}(GG|GG)} & {\scriptsize{}$8.942\times10^{-1}$} & {\scriptsize{}$5.624\times10^{-1}$} & {\scriptsize{}$5.218\times10^{-1}$} & {\scriptsize{}$8.378\times10^{-1}$} & {\scriptsize{}$4.581\times10^{-1}$} & {\scriptsize{}$2.033\times10^{-4}$} & {\scriptsize{}$1.493\times10^{-4}$}\tabularnewline
\hline 
\hline 
{\scriptsize{}total} & {\scriptsize{}$2.385\times10^{1}$} & {\scriptsize{}$3.592\times10^{0}$} & {\scriptsize{}$2.577\times10^{0}$} & {\scriptsize{}$1.225\times10^{1}$} & {\scriptsize{}$1.866\times10^{0}$} & {\scriptsize{}$8.519\times10^{-4}$} & {\scriptsize{}$7.843\times10^{-4}$}\tabularnewline
\end{tabular}
\par\end{centering}
\caption{Error $\Delta$ in the diagonal repulsion integrals $(\mu\nu|\mu\nu)$
in Hartree for the Fe atom in the def2-QZVP basis set employing the
fully uncontracted universal jfit\citep{Weigend2006_PCCP_65} and
jkfit\citep{Weigend2008_JCC_167} basis sets, the auxiliary basis
for MP2 calculations of the def2-QZVPP orbital basis set (def2-QZVPP-RIFIT),\citep{Haettig2005_PCCP_59}
as well as automatically generated sets using the AutoAbs\citep{Yang2007_JCP_74102}
and AutoAux\citep{Stoychev2017_JCTC_554} methods as well as the full
and reduced schemes of this work.\label{tab:diagonal-error}}
\end{sidewaystable*}

\subsection{Accuracy on the W4-17 database \label{subsec:Accuracy-on-the}}

\begin{figure}
\subfloat[2ZaPa-NR]{\begin{centering}
\includegraphics[width=0.5\textwidth]{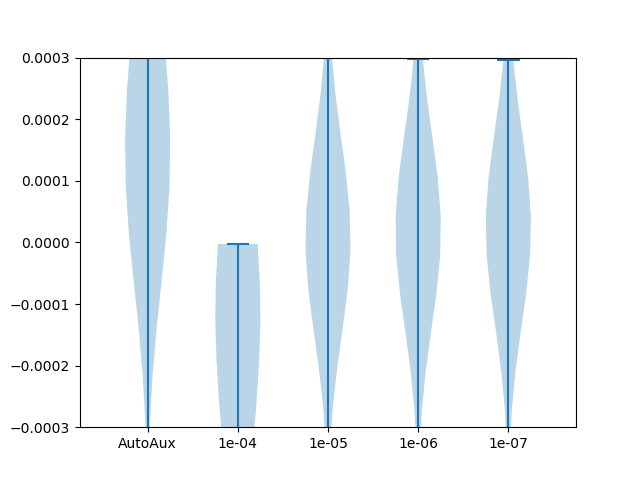}
\par\end{centering}

}\subfloat[3ZaPa-NR]{\begin{centering}
\includegraphics[width=0.5\textwidth]{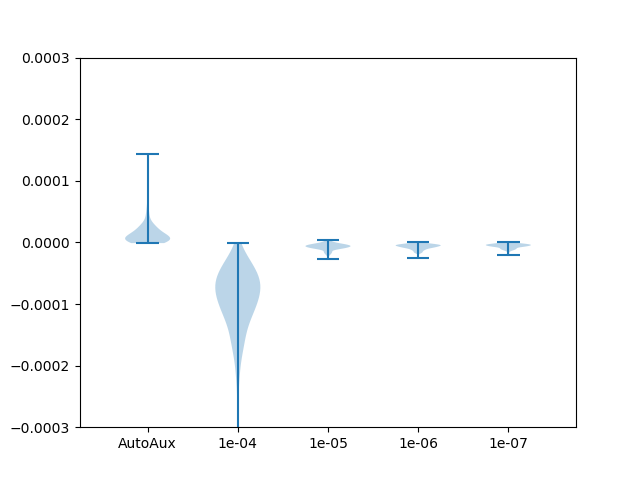}
\par\end{centering}
}

\subfloat[4ZaPa-NR]{\begin{centering}
\includegraphics[width=0.5\textwidth]{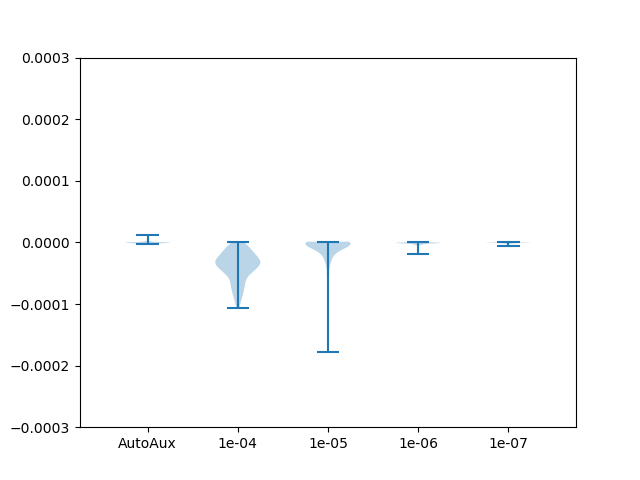}
\par\end{centering}
}\subfloat[5ZaPa-NR]{\begin{centering}
\includegraphics[width=0.5\textwidth]{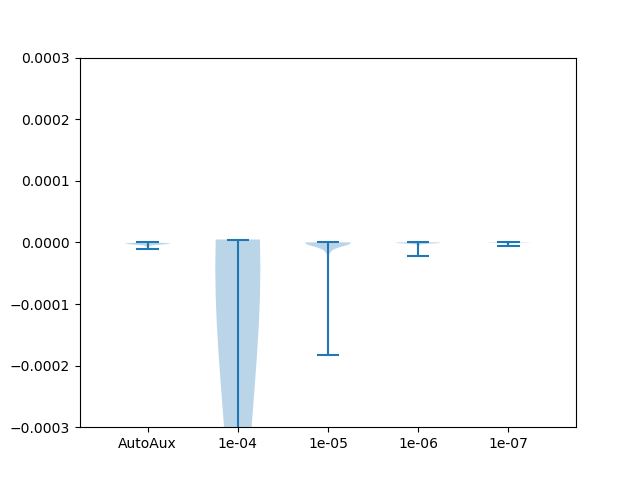}
\par\end{centering}
}\caption{RI errors in $E_{h}$ in the RI-HF total energy of the W4-17 dataset
for the AutoAux method and the reduced auxiliary basis obtained with
$\tau=10^{-4}$, $\tau=10^{-5}$, $\tau=10^{-6}$, and $\tau=10^{-7}$.
\label{fig:RIHF}}
\end{figure}

\begin{figure}
\subfloat[2ZaPa-NR]{\begin{centering}
\includegraphics[width=0.5\textwidth]{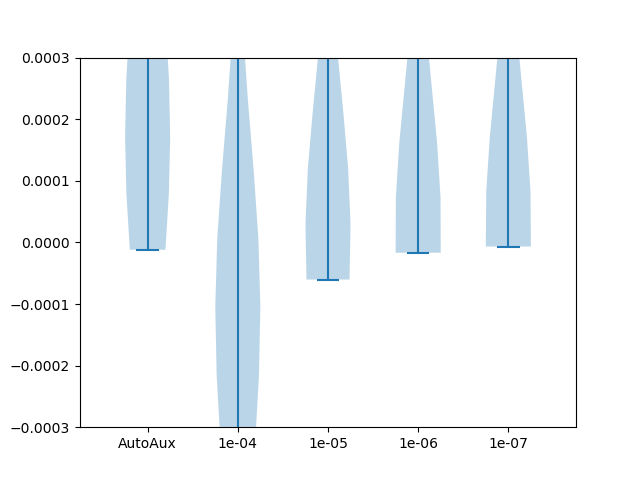}
\par\end{centering}
}\subfloat[3ZaPa-NR]{\begin{centering}
\includegraphics[width=0.5\textwidth]{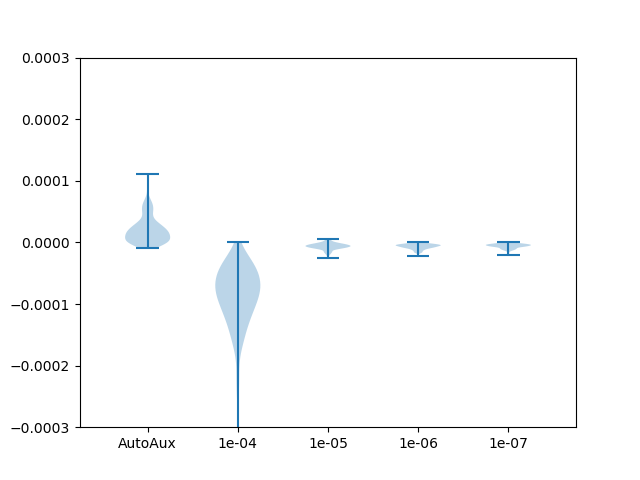}
\par\end{centering}
}

\subfloat[4ZaPa-NR]{\begin{centering}
\includegraphics[width=0.5\textwidth]{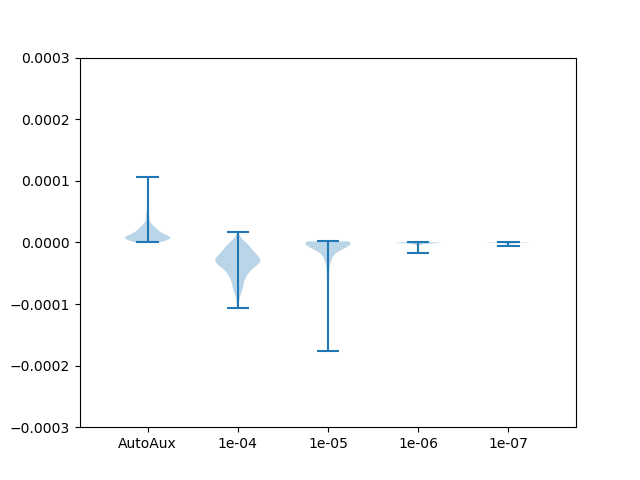}
\par\end{centering}
}\subfloat[5ZaPa-NR]{\begin{centering}
\includegraphics[width=0.5\textwidth]{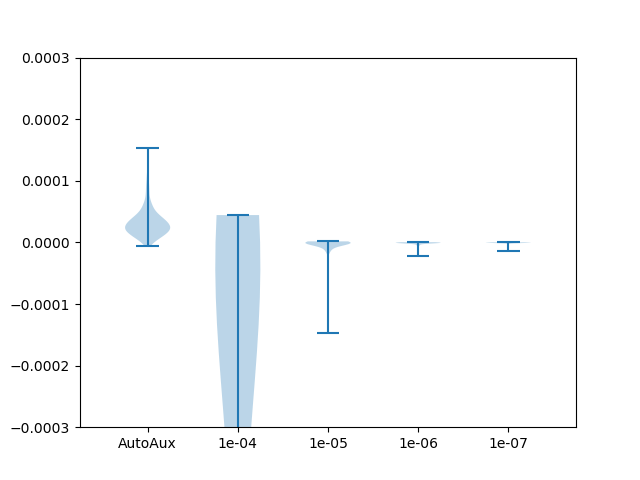}
\par\end{centering}
}\caption{RI errors in $E_{h}$ in the RI-MP2 total energy of the W4-17 dataset
for the AutoAux method and the reduced auxiliary basis obtained with
$\tau=10^{-4}$, $\tau=10^{-5}$, $\tau=10^{-6}$, and $\tau=10^{-7}$.\label{fig:RIMP2}}
\end{figure}

Violin plots\citep{Hintze1998_AS_181} whose width demonstrates the
distribution of the RI errors in the total energy at the RI-HF and
RI-MP2 levels of theory are shown in \figref{RIHF, RIMP2}, respectively.
The largest RI-HF and RI-MP2 errors encountered in the database are
shown in \tabref{autoaux, redbasis} for the AutoAux method of \citet{Stoychev2017_JCTC_554}
and the reduced auxiliary basis method of this work with $\tau=10^{-7}$,
respectively.

The errors in the range of millihartrees for the double-$\zeta$ 2ZaPa-NR
basis set are unacceptably large. The reason for such large errors
is that this orbital basis does not include sufficiently many high-angular-momentum
functions, which are necessary in the auxiliary basis set in order
to describe products of orbital basis functions located on different
atoms. This behavior is well-known also in the case of other automated
approaches for auxiliary basis set generation. If an auxiliary basis
set is needed for a small orbital basis set, an auxiliary basis generated
for a larger orbital basis set should be used instead to guarantee
sufficiently small RI errors.

The RI errors in the total energies in both RI-HF and RI-MP2 calculations
for triple-$\zeta$ and larger basis sets are small, ranging only
up to tens of microhartree as shown by \tabref{redbasis}, while the
AutoAux method yields somewhat larger errors in total energies as
shown by \tabref{autoaux}. However, even the largest error of $1.53\times10^{-4}\ E_{h}$
in the RI-MP2/5ZaPa-NR total energies with the AutoAux method is less
than 0.1 kcal/mol. Moreover, as errors in total energies tend to cancel
out in applications, both AutoAux and the present algorithm with a
suitably small value for $\tau$ appear to be reliable ways to generate
auxiliary basis sets for applications.

\begin{table*}
\begin{centering}
\begin{tabular}{c|cc|cc}
orbital basis & RI-HF ($E_{h}$) & molecule & RI-MP2 ($E_{h}$) & molecule\tabularnewline
\hline 
\hline 
2ZaPa-NR & $5.67\times10^{-3}$ & c-hooo & $9.07\times10^{-3}$ & c-hooo\tabularnewline
3ZaPa-NR & $1.44\times10^{-4}$ & hclo4 & $1.12\times10^{-4}$ & c2cl6\tabularnewline
4ZaPa-NR & $1.18\times10^{-5}$ & hclo4 & $1.06\times10^{-4}$ & c2cl6\tabularnewline
5ZaPa-NR & $1.11\times10^{-5}$ & p4 & $1.53\times10^{-4}$ & c2cl6\tabularnewline
\end{tabular}
\par\end{centering}
\caption{Largest RI errors in the database employing the AutoAux method.\citep{Stoychev2017_JCTC_554}
\label{tab:autoaux}}

\end{table*}

\begin{table*}
\begin{centering}
\begin{tabular}{c|cccc}
orbital basis & RI-HF ($E_{h}$) & molecule & RI-MP2 ($E_{h}$) & molecule\tabularnewline
\hline 
\hline 
2ZaPa-NR & $5.97\times10^{-3}$ & c-hooo & $8.82\times10^{-3}$ & c-hooo\tabularnewline
3ZaPa-NR & $2.00\times10^{-5}$ & c2cl6 & $1.97\times10^{-5}$ & c2cl6\tabularnewline
4ZaPa-NR & $5.37\times10^{-6}$ & dithiotane & $5.25\times10^{-6}$ & dithiotane\tabularnewline
5ZaPa-NR & $6.65\times10^{-6}$ & c2cl6 & $1.40\times10^{-5}$ & benzene\tabularnewline
\end{tabular}
\par\end{centering}
\caption{Largest RI errors in the database employing the reduced auxiliary
basis method of this work with $\tau=10^{-7}$. \label{tab:redbasis}}
\end{table*}

\subsection{Timing benchmark \label{subsec:Timing-benchmark}}

To further illustrate the usefulness of the RI approach and automatically
generated auxiliary basis sets, we consider calculations on commodity
hardware with free and open source software whose advantages for computational
chemistry has recently been discussed in \citeref{Lehtola2021__}.
As a practical example, we consider HF and MP2 calculations on small
water clusters with tight-binding geometries from \citeref{Miro2013_PCCP_1837}.
The 3ZaPa-NR orbital basis set\citep{Ranasinghe2013_JCP_144104} is
used, and the calculations are carried out with the \textsc{Psi4}
program\citep{Smith2020_JCP_184108} with various auxiliary basis
sets: in addition to sets generated using the algorithm of this work
with various values for $\tau$, we also consider the AutoAux algorithm.
Because the present algorithm employs uncontracted basis sets, data
are also included for an AutoAux basis generated for the 3ZaPa-NR
basis in fully uncontracted form (denoted as AutoAux{*}).

Before reporting the employed methodology and the resulting timings,
it is imperative to discuss the general limitations of published benchmarks
timings, which are well-known to be a sensitive topic in quantum chemistry:
in fact, some programs go as far as to forbid reporting performance
data altogether because of the issues discussed below. 

Obviously, the timings depend on the employed hardware. The present
calculations are performed on a commodity cloud server with an Intel
Xeon W3520 processor running at 2.67GHz, with 16 GB of memory. The
timings also depend the compiler as well as the employed compiler
options, linear algebra and two-electron integral libraries, and so
on. For simplicity, we employ the Fedora package of \textsc{Psi4}
version 1.3.2 on Fedora 34 (\texttt{psi4-1.3.2-10.fc34.x86\_64}),
and have not attempted to tune its performance in any way. (Note that
the RI calculations reported in \subsecref{Accuracy-on-the} were
also performed on the same hardware and program package.)

An even bigger impact on timings arise from the used program. The
employed algorithm has a huge impact: for instance, the evaluation
of the MP2 correlation energy by direct methods\citep{HeadGordon1988_CPL_503}
becomes highly appealing for large systems, but such an algorithm
is not available in \textsc{Psi4} at present. Different programs may
also employ different numerical thresholds by default, meaning they
are actually performing different types of calculations. All of these
issues should be kept in mind when examining the timing data presented
below, whose only purpose is to illustrate considerations in running
practical computations with \textsc{Psi4}.

Disk-based algorithms (\texttt{scf\_type pk} and \texttt{mp2\_type
conv}) were used for the conventional calculations, whereas the RI
calculations employed \texttt{scf\_type df} and \texttt{mp2\_type
df}; also other types of SCF and MP2 algorithms are available in \textsc{Psi4},
but they were not considered for the present benchmark. The calculations
employ 4 threads, 14 GB of memory, and the default convergence and
accuracy thresholds in \textsc{Psi4}. The resulting wall times are
shown in \tabref{Wall-times-for}, while the binding energies and
RI errors therein are shown in \tabref{RI-errors-in}.

As can be seen from \tabref{Wall-times-for}, while the conventional
algorithm is slightly faster for the smallest systems, there is a
crossover whose position depends on the employed auxiliary basis,
after which the RI calculations become significantly faster than the
conventional algorithm. While the conventional MP2 calculation for
the two-molecule cluster still employs an in-core algorithm, the MP2
step of the three-molecule cluster employs a partly out-of-core algorithm,
resulting in the considerable wall time difference between the HF
and MP2 calculations for this system. An even larger jump in the computational
cost of the conventional algorithm is observed for both HF and MP2
calculations on the four-molecule cluster, for which the two-electron
integrals no longer fit in memory. At variance, the RI integrals fit
in memory even for the largest computations considered, and no sudden
jumps in the computational cost are observed for the RI calculations.

The AutoAux algorithm, which has a large number of parameters that
have been optimized to yield the smallest possible auxiliary basis
sets that still are accurate enough for routine applications, is competitive
with the conventional algorithm even for the smallest system size
studied, the water molecule. The AutoAux algorithm replaces contracted
Gaussian functions with an effective Gaussian primitive. Comparison
of the timings for the AutoAux auxiliary basis sets generated for
the 3ZaPa-NR basis in its original contracted and fully uncontracted
form shows that this design aspect of the AutoAux algorithm has a
significant effect on its computational performance: the calculations
that employ an auxiliary basis generated for the 3ZaPa-NR basis in
uncontracted form are 2--3 times slower.

Interestingly, as can be seen by studying the data in \figref{RIHF, RIMP2}
and \tabref{Wall-times-for, RI-errors-in}, the accuracy and computational
performance of the present algorithm with $\tau=10^{-5}$ appears
to be roughly comparable to AutoAux for uncontracted basis sets, even
though AutoAux is a heavily optimized semiempirical algorithm with
tens of parameters that also truncates the high-angular-momentum functions,
while the present algorithm is controlled by a single parameter, allowing
an easy adjustment of the computational cost / accuracy ratio, and
no further truncation of high-angular-momentum functions has been
attempted in this work. Moreover, the present algorithm affords more
accurate results (with respect to the complete auxiliary basis set
limit) at only slightly increased computational cost by the use of
a smaller value of $\tau$, such as $\tau=10^{-6}$. These results
are highly promising for future applications of the present algorithm
to Slater-type orbital and numerical atomic orbital basis sets: the
auxiliary basis sets generated by the present algorithm are competitive
with ones from the heavily optimized AutoAux algorithm when uncontracted
basis sets are employed, proving the expected optimal nature of the
present algorithm.

\begin{table*}
\subfloat[HF]{\begin{centering}
\begin{tabular}{lrrrrrrrr}
 & $\tau=10^{-3}$ & $\tau=10^{-4}$ & $\tau=10^{-5}$ & $\tau=10^{-6}$ & $\tau=10^{-7}$ & AutoAux{*} & AutoAux & conv\tabularnewline
\hline 
\hline 
\ce{H2O} & 3.9 & 4.1 & 4.2 & 4.4 & 4.5 & 4.2 & 3.8 & 3.8\tabularnewline
\ce{(H2O)2} & 6.4 & 8.0 & 9.6 & 10.8 & 12.0 & 9.2 & 5.9 & 9.7\tabularnewline
\ce{(H2O)3} & 13.9 & 20.1 & 26.0 & 31.4 & 36.5 & 25.9 & 12.4 & 34.9\tabularnewline
\ce{(H2O)4} & 29.1 & 44.0 & 59.7 & 72.1 & 103.5 & 56.4 & 25.2 & 1492.5\tabularnewline
\ce{(H2O)5} & 80.5 & 127.1 & 160.6 & 184.1 & 212.6 & 151.2 & 51.3 & 3951.3\tabularnewline
\end{tabular}
\par\end{centering}
}

\subfloat[MP2]{\begin{centering}
\begin{tabular}{lrrrrrrrr}
 & $\tau=10^{-3}$ & $\tau=10^{-4}$ & $\tau=10^{-5}$ & $\tau=10^{-6}$ & $\tau=10^{-7}$ & AutoAux{*} & AutoAux & conv\tabularnewline
\hline 
\hline 
\ce{H2O} & 5.0 & 5.4 & 8.9 & 5.9 & 6.2 & 5.5 & 4.7 & 4.6\tabularnewline
\ce{(H2O)2} & 8.8 & 11.6 & 14.0 & 16.3 & 18.4 & 13.6 & 8.1 & 27.8\tabularnewline
\ce{(H2O)3} & 20.4 & 30.2 & 40.4 & 48.9 & 57.6 & 39.7 & 18.0 & 221.7\tabularnewline
\ce{(H2O)4} & 43.3 & 67.2 & 93.1 & 112.8 & 152.1 & 86.0 & 36.8 & 2408.5\tabularnewline
\ce{(H2O)5} & 115.3 & 178.6 & 233.4 & 268.0 & 314.0 & 212.3 & 76.7 & 7424.0\tabularnewline
\end{tabular}
\par\end{centering}

}

\caption{Total wall times in seconds for HF and MP2 calculations on water clusters
in the 3ZaPa-NR orbital basis, using either an RI algorithm with automatically
generated basis sets with the present algorithm with various values
for $\tau$ or the AutoAux method,\citep{Stoychev2017_JCTC_554} or
the conventional algorithm (conv). AutoAux{*} denotes an AutoAux basis
generated for an uncontracted orbital basis. The timings for the MP2
calculations also include the HF step.\label{tab:Wall-times-for}}

\end{table*}

\begin{table*}
\subfloat[HF]{\begin{centering}
\begin{tabular}{llrrrrrrr}
 & $E_{\text{bind}}^{\text{conventional}}$ & $\tau=10^{-3}$ & $\tau=10^{-4}$ & $\tau=10^{-5}$ & $\tau=10^{-6}$ & $\tau=10^{-7}$ & AutoAux{*} & AutoAux\tabularnewline
\hline 
\hline 
\ce{(H2O)2} & 0.130 & 0.051 & 0.065 & -0.000 & 0.004 & 0.003 & -0.001 & -0.001\tabularnewline
\ce{(H2O)3} & 0.372 & 0.084 & 0.094 & -0.004 & 0.032 & 0.021 & -0.010 & -0.001\tabularnewline
\ce{(H2O)4} & 0.703 & 0.107 & 0.131 & -0.004 & 0.041 & 0.028 & -0.007 & 0.004\tabularnewline
\ce{(H2O)5} & 0.971 & 0.153 & 0.164 & -0.005 & 0.046 & 0.042 & -0.009 & 0.006\tabularnewline
\end{tabular}
\par\end{centering}
}

\subfloat[MP2]{\begin{centering}
\begin{tabular}{llrrrrrrr}
 & $E_{\text{bind}}^{\text{conventional}}$ & $\tau=10^{-3}$ & $\tau=10^{-4}$ & $\tau=10^{-5}$ & $\tau=10^{-6}$ & $\tau=10^{-7}$ & AutoAux{*} & AutoAux\tabularnewline
\hline 
\hline 
\ce{(H2O)2} & 0.216 & 0.097 & 0.073 & 0.000 & 0.004 & 0.003 & -0.011 & -0.009\tabularnewline
\ce{(H2O)3} & 0.662 & 0.190 & 0.110 & -0.004 & 0.031 & 0.020 & -0.042 & -0.028\tabularnewline
\ce{(H2O)4} & 1.204 & 0.267 & 0.158 & -0.003 & 0.039 & 0.025 & -0.076 & -0.058\tabularnewline
\ce{(H2O)5} & 1.580 & 0.353 & 0.200 & -0.003 & 0.044 & 0.039 & -0.105 & -0.083\tabularnewline
\end{tabular}
\par\end{centering}
}

\caption{HF and MP2 binding energies $E_{\text{bind}}(n)=nE[\text{H}_{2}\text{O}]-E[(\text{H}_{2}\text{O})_{n}]$
of water clusters in eV, as well as the corresponding RI errors $\Delta E_{\text{bind}}(n)=E_{\text{bind}}^{\text{conventional}}(n)-E_{\text{bind}}^{\text{RI}}(n)$
for various automatically generated auxiliary basis sets in meV. All
calculations employ the 3ZaPa-NR orbital basis set and tight-binding
geometries from \citeref{Miro2013_PCCP_1837}. The notation is otherwise
analogous to that in \tabref{Wall-times-for}. \label{tab:RI-errors-in}}
\end{table*}

\subsection{Importance of preordering: case study \label{subsec:Importance-of-preordering}}

As a final point, we will investigate the importance of the preordering
of the auxiliary function candidates before the pivoted Cholesky decomposition.
As the ordering presumably becomes more important for larger basis
sets, we will examine the 5ZaPa-NR orbital basis for this part of
the study, and employ the full algorithm to maximize the number of
candidates.

\begin{figure}
\begin{centering}
\includegraphics{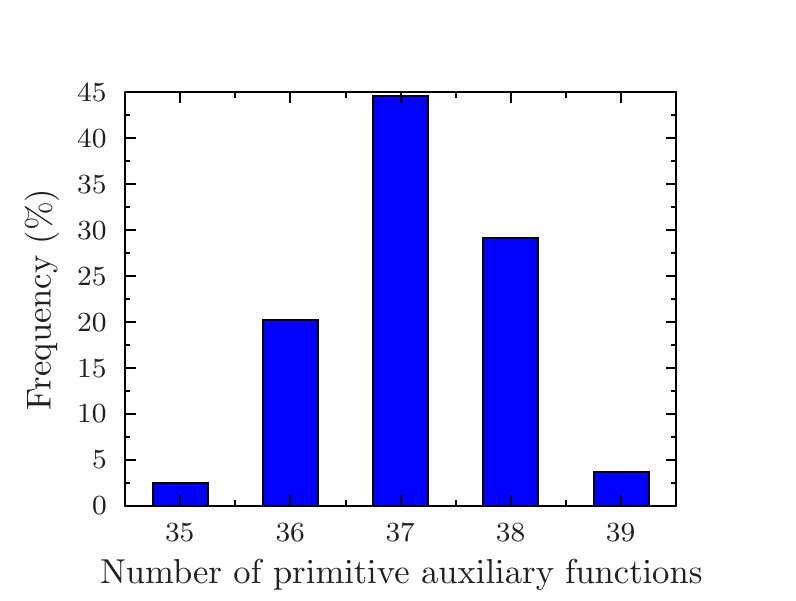}
\par\end{centering}
\caption{Distribution of the number of P primitives in automatically generated
auxiliary basis sets for the 5ZaPa-NR orbital basis for Ar, sampled
for 10 000 random permutations of the 818 candidate functions. \label{fig:Number-of-P}}

\end{figure}

We begin by examining the P shell of auxiliary functions for Ar, for
which the full algorithm leads to 818 primitive candidate functions.
Performing 10 000 random permutations of the 818 primitives and running
the pivoted Cholesky algorithm on each of them, we obtain the distribution
shown in \figref{Number-of-P}. In contrast, feeding in the candidates
in numerical order of the exponents, or ordering the candidate functions
in increasing off-diagonal overlap both result in 37 primitives---which
coincides with the maximum of the distribution in \figref{Number-of-P}.
Thus, even though some of the randomized initial orderings resulted
in a slightly smaller number of auxiliary functions, the level of
performance for the presently employed algorithm is satisfactory.

Moving on to the comparison of the compositions of the complete auxiliary
basis sets for the 5ZaPa-NR orbital basis resulting from ordering
the candidates on the individual shells of the individual elements
H--Ar either by i) numerical value or ii) off-diagonal Coulomb overlap,
only small differences are observed (not shown): most shells end up
with the same number of functions, while the others differ only by
the addition or removal of a single function. Although mixed performance
is observed for methods i) and ii), with method i) sometimes yielding
fewer functions than method ii), method ii) yields the fewest number
of functions overall. Future implementations of the present algorithm
may attempt to minimize the size of the generated auxiliary basis
by trying out several initial orderings of the candidate functions
before the pivoted Cholesky decomposition, and picking the one that
leads to the most compact decomposition i.e. smallest auxiliary basis
set for each shell.

\section{Summary and discussion\label{sec:Summary}}

We have presented a general method to form auxiliary basis sets in
linear combination of atomic orbitals calculations. In addition to
commonly-used Gaussian basis sets, the method is also straightforwardly
applicable to use with Slater-type orbitals and numerical atomic orbitals,
and we hope to pursue such applications in future work. The algorithm
can be implemented in a simple-minded manner from the full set of
orbital products; alternatively, a Cholesky decomposition of the atomic
two-electron integrals tensor can be used to screen the set of candidate
functions, leading to auxiliary basis sets with considerably fewer
functions at intermediate angular momenta.

We have benchmarked the accuracy of the reduced auxiliary basis sets
in proof-of-concept applications to Hartree--Fock and second-order
Møller--Plesset perturbation theory calculations on the non-multireference
part of the W4-17 database of molecules, and shown that the resulting
fitting errors in total energies are negligible for basis sets of
at least polarized triple-$\zeta$ quality. The Gaussian auxiliary
basis sets obtained with the present algorithm compare favorably in
accuracy and computational cost to the heavily parametrized AutoAux
method\citep{Stoychev2017_JCTC_554} when the orbital basis is not
contracted. These results are promising for the future applications
with Slater-type and numerical atomic orbital basis sets, for which
the computation of the exact two-electron integrals is difficult.

As the full auxiliary sets are larger than the reduced auxiliary basis
sets, they are expected to be at least as accurate as the reduced
auxiliary basis sets. Therefore they should likewise be useful in
applications where Cholesky decompositions of the atomic two-electron
integrals tensor are not available. The full auxiliary basis set algorithm,
which was presented in \algref{Flowchart-algorithm.}, is similar
to the one of \citet{Ren2012_NJP_53020}; the main difference being
that their Gram--Schmidt procedure is replaced by the pivoted Cholesky
decomposition. Thus, implementing the present method in existing implementations
of the \citeauthor{Ren2012_NJP_53020} algorithm should be extremely
straightforward. (As a further difference, we also suggest reordering
the auxiliary function candidates in increasing off-diagonal overlap.)

The automated schemes pursued in this work as well as \citerefs{Ren2012_NJP_53020}
and \citenum{Aquilante2009_JCP_154107} do not truncate high-angular-momentum
functions unlike e.g. the schemes of \citet{Yang2007_JCP_74102} and
\citet{Stoychev2017_JCTC_554} Although the high-angular-momentum
functions can usually be discarded for global Coulomb and exchange
fitting, the situation is different for e.g. local fitting algorithms
that may be used to circumvent the steep computational scaling of
global fitting methods.\citep{Ihrig2015_NJP_93020} Possible strategies
for further truncation of the generated auxiliary basis sets may be
investigated in future work.

Because the procedure of this work is able to generate auxiliary basis
sets of varying precision, and because the auxiliary basis set arising
from the decomposition of \eqref{coulovl} to threshold $\tau_{1}$
is a subset of the one obtained for $\tau_{2}<\tau_{1}$, the present
algorithm could be easily paired with e.g. dual auxiliary basis set
approaches,\citep{Csoka2021_JCP_164114} in analogy to the established
practice of employing Cholesky decompositions of the two-electron
integrals of varying thresholds\citep{Aquilante2007_JCP_194106} to
speed up self-consistent field calculations.

As was already discussed in \citerefs{Lehtola2019_JCP_241102} and
\citenum{Lehtola2020_PRA_32504}, the pivoted Cholesky procedure depends
somewhat on the initial ordering of the basis functions, as the (Coulomb)
overlap matrix has a unit diagonal. This work used the self-sufficient
method introduced in \citeref{Lehtola2020_PRA_32504} of ordering
the functions in increasing off-diagonal overlap. But, it was also
found that slightly more compact decompositions, that is, auxiliary
basis sets can be found in some cases by a randomized search. As generating
the auxiliary basis is inexpensive compared to the subsequent electronic
structure calculations, future implementations of the present algorithm
may consider employing quasi-exhaustive searches to minimize the size
of the generated auxiliary basis set, even though only small differences
are expected in the resulting numbers of primitive functions.

In this work, we have employed the pivoted Cholesky procedure of \citerefs{Lehtola2019_JCP_241102}
and \citenum{Lehtola2020_PRA_32504} to choose the auxiliary functions
from a pool of candidate functions. As our last point, we wish to
note that the same algorithm can also be used to choose linearly independent
sets of auxiliary functions in \emph{molecular} \emph{calculations}.
Overcompleteness issues with the auxiliary basis set are expected
in molecular calculations with extended basis sets such as those developed
in \citeref{Lehtola2020_JCP_134108}; the modeling of weakly bound
electrons\citep{Herbert2015__391} also often involves extremely diffuse
basis functions.\citep{Lehtola2019_JCP_241102} The density-fitted
two-electron integrals of \eqref{RI} can be written as $(\mu\nu|\sigma\rho)=\sum_{A}B_{\mu\nu}^{A}B_{\sigma\rho}^{A}$,
where the ``B matrix'' is defined as $B_{\mu\nu}^{A}=\sum_{B}(A|B)^{-1/2}(B|\mu\nu)$.
The formation of $(A|B)^{-1/2}$ is analogous to the canonical orthogonalization
algorithm for orbital basis sets,\citep{Loewdin1956_AP_1} and may
be numerically ill-conditioned. However, continuing the analogy, $(A|B)^{-1/2}$
can be formed in a numerically stable manner by employing the procedure
of \citerefs{Lehtola2019_JCP_241102} and \citenum{Lehtola2020_PRA_32504}:
a pivoted Cholesky decomposition on $S_{AB}=(A|B)$ can be used to
pick a linearly independent subset of auxiliary functions for the
molecule, which can then be employed to carry out the calculation.

\section*{Appendix I. Coulomb overlap integrals}

As is well known,\citep{Lehtola2019_IJQC_25945,Lehtola2019_IJQC_25968}
one-center two-electron integrals reduce to a product of radial and
angular integrals through the Laplace expansion
\[
\frac{1}{r_{12}}=\frac{4\pi}{r_{>}}\sum_{L=0}^{\infty}\frac{1}{2L+1}\left(\frac{r_{<}}{r_{>}}\right)^{L}\sum_{M=-L}^{L}Y_{L}^{M}(\Omega_{1})\left(Y_{L}^{M}(\Omega_{2})\right)^{*}.
\]
The case of four-index integrals leads to the need to evaluate Gaunt
coefficients and various types of integrals; see \citeref{Lehtola2019_IJQC_25945}
for discussion. However, as the present scheme only requires two-index
integrals, only simple radial integrals are necessary due to the orthonormality
of the spherical harmonics
\begin{equation}
(A|B)=\int\frac{\chi_{A}^{*}(\boldsymbol{r})\chi_{B}(\boldsymbol{r}')}{\left|\boldsymbol{r}-\boldsymbol{r}'\right|}{\rm d}^{3}r\,{\rm d}^{3}r'=\frac{\delta_{l_{A}l_{B}}}{2l_{A}+1}\int_{0}^{\infty}{\rm d}rr^{2}\int_{0}^{\infty}dr'r^{2}\frac{4\pi}{r_{>}}\left(\frac{r_{<}}{r_{>}}\right)^{l_{A}}\chi_{A}(r)\chi_{B}(r')\label{eq:coulovl-def}
\end{equation}
The radial integral can be rewritten in a form more suitable for computation
as
\begin{align*}
(A|B) & =\iint\chi_{A}(r)\chi_{B}(r')\frac{r_{<}^{l}}{r_{>}^{l+1}}r^{2}{\rm d}rr'^{2}{\rm d}r'\\
 & =\int_{0}^{\infty}{\rm d}rr^{2}\int_{0}^{r}{\rm d}r'r'^{2}\left[f(r)g(r')+f(r')g(r)\right]\frac{r'^{l}}{r^{l+1}}\\
 & =\int_{0}^{\infty}{\rm d}rr^{2-l-1}\int_{0}^{r}{\rm d}r'r'^{2+l}\left[f(r)g(r')+f(r')g(r)\right]
\end{align*}

\subsection*{Gaussian type orbitals}

Spherical harmonic Gaussian type orbitals (GTOs) have the form
\[
\chi_{A}^{\text{GTO}}(r)=r^{l_{A}}e^{-\alpha_{A}r^{2}}Y_{l_{A}}^{m_{A}}(\hat{\boldsymbol{r}})
\]
which yields
\begin{align*}
(A|B)^{\text{GTO}}= & \int_{0}^{\infty}{\rm d}rr^{2-l-1}\int_{0}^{r}{\rm d}r'r'^{2+l}\left[f(r)g(r')+f(r')g(r)\right]\delta_{ll_{A}}\delta_{ll_{B}}\\
= & \int_{0}^{\infty}{\rm d}rr\int_{0}^{r}{\rm d}r'r'^{2+2l}\left[e^{-\alpha_{A}r^{2}-\alpha_{B}r'^{2}}+e^{-\alpha_{A}r'^{2}-\alpha_{B}r{}^{2}}\right]\delta_{ll_{A}}\delta_{ll_{B}}\\
= & \frac{1}{4}\Gamma\left(l+\frac{3}{2}\right)\frac{(\alpha_{A}+\alpha_{B})^{-l-1/2}}{\alpha_{A}\alpha_{B}}\delta_{ll_{A}}\delta_{ll_{B}}
\end{align*}
with Mathematica 12.1.

\subsection*{General Slater type orbitals}

Slater type orbitals (STOs) have the general form
\[
\chi_{A}^{\text{STO}}(r)=r^{n_{A}-1}e^{-\alpha_{A}r}Y_{l_{A}}^{m_{A}}(\hat{\boldsymbol{r}})
\]
which yields
\begin{align*}
 & \int_{0}^{\infty}{\rm d}rr^{1-l}\int_{0}^{r}{\rm d}r'r'^{2+l}\left[f(r)g(r')+f(r')g(r)\right]\\
= & \int_{0}^{\infty}{\rm d}rr^{-l}\int_{0}^{r}{\rm d}r'r'^{l+1}\left[r^{n_{A}}r'^{n_{B}}e^{-\alpha_{A}r^{2}-\alpha_{B}r'^{2}}+r^{n_{B}}r'^{n_{A}}e^{-\alpha_{A}r'^{2}-\alpha_{B}r{}^{2}}\right]
\end{align*}
Evaluation with Mathematica 12.1 leads to
\begin{align*}
 & \alpha_{A}^{-2-L-n_{A}}\alpha_{B}^{-1+L-n_{B}}\Gamma(2+L+n_{A})\Gamma(1-L+n_{B})\\
+ & \alpha_{A}^{-1+L-n_{A}}\alpha_{B}^{-2-L-n_{B}}\Gamma(1-L+n_{A})\Gamma(2+L+n_{B})\\
- & \Gamma(3+n_{A}+n_{B})\times\Bigg[\\
 & \frac{\alpha_{B}^{-3-n_{A}-n_{B}}\ _{2}F_{1}\left(1-L+n_{A},3+n_{A}+n_{B},2-L+n_{A},-\frac{\alpha_{A}}{\alpha_{B}}\right)}{1-L+n_{A}}\\
 & \frac{\alpha_{A}^{-3-n_{A}-n_{B}}\ _{2}F_{1}\left(1-L+n_{B},3+n_{A}+n_{B},2-L+n_{B},-\frac{\alpha_{B}}{\alpha_{A}}\right)}{1-L+n_{B}}\Bigg]
\end{align*}
where $_{2}F_{1}$ is the ordinary hypergeometric function.

\subsection*{Even-tempered Slater type orbitals}

Even-tempered STOs pick the lowest possible primary quantum number
for all angular momentum channels, i.e. $n_{A}=l_{A}+1$, so that
the radial functions become analogous to GTOs 
\[
\chi_{A}^{\text{ET-STO}}(r)=r^{l_{A}}e^{-\zeta_{A}r}Y_{l_{A}}^{m_{A}}(\hat{\boldsymbol{r}})
\]
yielding a simpler expression
\[
(A|B)^{\text{ET-STO}}=\frac{[\zeta_{A}^{2}+\zeta_{B}^{2}+\zeta_{A}\zeta_{B}(3+2l)]\Gamma(3+2l)}{\zeta_{A}^{2}\zeta_{B}^{2}(\zeta_{A}+\zeta_{B})^{3+2l}}\delta_{ll_{A}}\delta_{ll_{B}}
\]

\section*{Appendix II. Effective Gaussian exponents}

The radial expectation value for a given radial function $R(r)=r^{l}e^{-\alpha r^{2}}$
is 
\begin{equation}
\langle r\rangle=\frac{\int_{0}^{\infty}r^{3}R(r)^{2}{\rm d}r}{\int_{0}^{\infty}r^{2}R(r)^{2}{\rm d}r}=\frac{\Gamma(l+2)}{\Gamma(l+\frac{3}{2})\sqrt{2\alpha}}.\label{eq:r-exp}
\end{equation}
This means that the contribution of a given product function $r^{l_{i}+l_{j}}e^{-(\alpha_{i}+\alpha_{j})r^{2}}$
into the angular momentum channel $|l_{i}-l_{j}|\leq L\leq l_{i}+l_{j}$
is best approximated by the function that satisfies
\begin{equation}
\frac{\Gamma(L+2)}{\Gamma(L+\frac{3}{2})\sqrt{2\alpha_{\text{eff}}}}=\frac{\Gamma(l_{i}+l_{j}+2)}{\Gamma(l_{i}+l_{j}+\frac{3}{2})\sqrt{2\left(\alpha_{i}+\alpha_{j}\right)}}\label{eq:rexp-equal}
\end{equation}
from which
\begin{equation}
\alpha_{\text{eff}}=\left[\frac{\Gamma(L+2)\Gamma(l_{i}+l_{j}+\frac{3}{2})}{\Gamma(l_{i}+l_{j}+2)\Gamma(L+\frac{3}{2})}\right]^{2}\left(\alpha_{i}+\alpha_{j}\right)\label{eq:a-eff}
\end{equation}
The gamma function scaling factor in \eqref{a-eff} is unity for $L=l_{i}+l_{j}$
and decreases monotonically for $L<l_{i}+l_{j}$ because in that case
$r^{l_{i}+l_{j}}$ results in a more diffuse character for the product
wave function than $r^{L}$ does, which is taken into account by scaling
down the exponent; for instance, for a $g\times g$ product coupling
to $L=0$ the scale factor is roughly 0.1375.

\subsection*{Acknowledgments}

We thank Frank Neese and Georgi Stoychev for assistance in verifying
the Basis Set Exchange implementation of the AutoAux method, Roland
Lindh for discussions on the acCD method, as well as Volker Blum for
discussions on local fitting methods. We thank the National Science
Foundation for financial support under Grant No. CHE-2136142.
\begin{tocentry}
\includegraphics{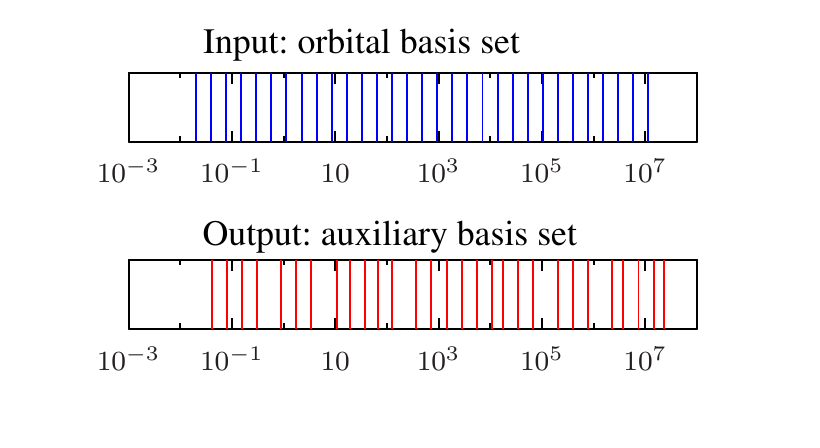}
\end{tocentry}
\bibliography{citations}

\end{document}